\documentclass[modern]{aastex}

\newcommand{\bvf}{Brunt-V\"ais\"al\"a }
\newcommand{\teff}{$\rm T_{eff}$}
\newcommand{\logg}{$\log$\ g }

\newcommand{\sigrms}{$\sigma_{\rm RMS}$}
\newcommand{\mstar}{$M_*$}
\newcommand{\menv}{$M_{\rm env}$}
\newcommand{\mhe}{$M_{\rm He}$}
\newcommand{\xo}{$X_o$}
\newcommand{\qfm}{$X_{\rm fm}$}
\newcommand{\muHz}{\mbox{$\mu$Hz}}
\newcommand{\msun}{$M_\odot$}

\shorttitle{Asteroseismology of GD358}
\shortauthors{Bischoff-Kim}
\usepackage{graphicx}
\usepackage{epsfig}
\usepackage[outdir=./]{epstopdf}
\usepackage{ulem}

\begin{document}   

\title{GD358: three decades of observations for the in-depth asteroseismology of 
a DBV star}

\author {Agn\`es~Bischoff-Kim\altaffilmark{1},
J.~L. Provencal\altaffilmark{2,3},
P.~A. Bradley\altaffilmark{6},
M.~H.~Montgomery\altaffilmark{3,4},
H.~L. Shipman\altaffilmark{2,3},
Samuel~T.~Harrold\altaffilmark{4},
B.~Howard\altaffilmark{3},
W.~Strickland\altaffilmark{5},
D.~Chandler\altaffilmark{5},
D.~Campbell\altaffilmark{5},
A.~Arredondo\altaffilmark{5},
R.~Linn\altaffilmark{5},
D.~P.~Russell\altaffilmark{5},
D.~Doyle\altaffilmark{5},
A.~Brickhouse\altaffilmark{5},
D.~Peters\altaffilmark{5},
S.-L. Kim\altaffilmark{7},
X.~J.~Jiang\altaffilmark{8},
Y-N.~Mao\altaffilmark{8},
A.~V.~Kusakin\altaffilmark{9},
A.~V.~Sergeev\altaffilmark{10,11},
M.~Andreev\altaffilmark{10,11},
S. Velichko\altaffilmark{10,11},
R.~Janulis\altaffilmark{12},
E.~Pakstiene\altaffilmark{12},
F.~Ali\c cavu\c s\altaffilmark{13},
N. Horoz\altaffilmark{13},
S.~Zola\altaffilmark{14,15},
W.~Og{\l}oza\altaffilmark{15},
D.~Koziel-Wierzbowska\altaffilmark{14},
T.~Kundera\altaffilmark{14},
D. Jableka\altaffilmark{14},
B. Debski\altaffilmark{14},
A. Baran\altaffilmark{15},
S. Meingast\altaffilmark{16},
T. Nagel\altaffilmark{17},
L. Loebling\altaffilmark{17},
C. Heinitz\altaffilmark{17},
D. Hoyer\altaffilmark{17},
Zs. Bogn\'ar\altaffilmark{18},
B.~G.~Castanheira\altaffilmark{19}
}
\altaffiltext{1}{Penn State Scranton, Dunmore, PA 18512; 
axk55@psu.edu}
\altaffiltext{2}{University of Delaware, Department of Physics and Astronomy
Newark, DE 19716; jlp@udel.edu}
\altaffiltext{3}{Delaware Asteroseismic Research Center, Mt. Cuba Observatory,
Greenville, DE 19807}
\altaffiltext{4}{Department of Astronomy, University of Texas, Austin, TX 78712;
mikemon@rocky.as.utexas.edu}
\altaffiltext{5}{Meyer Observatory and Central Texas Astronomical Society, 209
Paintbrush, Waco, TX 76705; chandler@vvm.com}
\altaffiltext{6}{Los Alamos National Laboratory, Los Alamos, NM 87545; 
pbradley@lanl.gov}
\altaffiltext{7}{Korea Astronomy and Space Science Institute, Daejeon 34055, 
Korea}
\altaffiltext{8}{National Astronomical Observatories, Academy of Sciences, 
Beijing 10012, People's Republic of
China; xjjiang@bao.ac.cn}
\altaffiltext{9}{Fesenkov Astrophysical Institute, Almaty 050020, Kazakhstan}
\altaffiltext{10}{Ukrainian National Academy of Sciences, International Center 
for Astronomical,
Medical and Ecological Research, 27, Akademika Zabolotnoho Ave., 03680 Kyiv, 
Ukraine; sergeev@terskol.com }
\altaffiltext{11}{Russian Academy of Sciences, Institute of Astronomy, Terskol 
Branch, 81, Elbrus Ave., ap. 33, Tyrnyauz,
Kabardino-Balkaria Republic, 361623, Russian Federation; sergeev@terskol.com}
\altaffiltext{12}{Institute of Theoretical Physics and Astronomy, Vilnius
University, Vilnius, Lithuania; jr@itpa.lt}
\altaffiltext{13}{Ulupinar Observatory, \c Canakkale Onsekiz Mart University, 
Turkey}
\altaffiltext{14}{Astronomical Observatory of the Jagiellonian University, ul. 
Orla
171, 30-244 Cracow, Poland; szola@oa.uj.edu.pl}
\altaffiltext{15}{Mount Suhora Observatory, Cracow Pedagogical University, Ul.
Podchorazych 2, 30-084 Krakow, Poland; zola@astro1.as.ap.krakow.pl}
\altaffiltext{16}{Institut f\H ur Astrophysik, Universit\H at Wien, T\H 
urkenschanzstrasse 17, 1180 Wien, Austria}
\altaffiltext{17}{Institut fuer Astronomie und Astrophysik, Kepler Center for 
Astro and Particle Physics,
Eberhard Karls Universitaet Tuebingen, Sand 1, 72076 Tuebingen Germany; 
astro-teleskop-owner@listserv.uni-tuebingen.de}
\altaffiltext{18}{Konkoly Observatory, MTA CSFK, Konkoly Thege M. u. 15-17, 
H-1121 Budapest, Hungary}
\altaffiltext{19}{Baylor University, Department of Physics, Waco, TX 76798}
\clearpage
\newpage

\begin{abstract}

We report on the analysis of 34 years of photometric observations of the 
pulsating helium atmosphere white dwarf GD358. The complete data set includes 
archival data from 1982-2006, and 1195.2 hours of new observations from 
2007-2016. From this data set, we extract 15 frequencies representing g-mode 
pulsation modes, adding 4 modes to the 11 modes known previously. We present 
evidence that these 15 modes are $\ell=1$ modes, 13 of which belong to a 
consecutive sequence in radial overtone $k$. We perform a detailed asteroseismic 
analysis using models that include parameterized, complex carbon and oxygen core 
composition profiles to fit the periods. Recent spectroscopic analyses place 
GD358 near the red edge of the DBV instability strip, at 24,000 $\pm$ 500~K and a 
\logg of 7.8 $\pm$ 0.08 dex. The surface gravity translates to a mass range of 
0.455 to 0.540 \msun . Our best fit model has a temperature of 23,650~K and a 
mass of 0.5706 \msun . That is slightly more massive than suggested by most the 
recent spectroscopy. We find a pure helium layer mass of $10^{-5.50}$, 
consistent with the result of previous studies and the outward diffusion of helium over time.

\end{abstract}


\section{Astrophysical Context}
\label{intro}
White dwarfs (WDs) represent the final phase of evolution for around 98\% of the stellar
population in our Galaxy. Concealed in their interior structure and composition are the 
fingerprints of physical processes that took place during earlier stages in their life 
cycles. For example, nuclear reaction rates during the progenitor's core helium burning 
phase fix the resulting white dwarf's core composition. The relative time spent burning 
hydrogen and helium during the progenitor's asymptotic-giant-branch (AGB) phase and 
accompanying mass-loss episodes determine the final white dwarf helium layer thickness 
\citep{Lawlor06,Althaus05}. White dwarfs come in two basic flavors depending on their 
surface layer composition, something which is again determined 
by processes occurring in the last stages of stellar evolution. Helium atmosphere 
white dwarfs (DBs) comprise roughly 20\% of the population of field white 
dwarfs, with most of the remaining 80\% consisting of their hydrogen atmosphere 
(DA) cousins. A leading theory behind the bifurcation into two spectral classes
involves a very late thermal pulse that burns off residual hydrogen in the envelope, 
producing a nearly pure helium atmosphere \citep{Iben83}. Such objects proceed along 
the white dwarf cooling track as PG 1159 stars, which are widely believed to be a
class of precursors of DB white dwarfs. DBs are known to pulsate at effective temperatures 
ranging between 21,000 K and 31,000 K (DBVs) \citep{Beauchamp99, Castanheira05, Hermes2017}. 

The subject of this paper, GD358 (V777 Her) is the brightest ($m_{\rm v}=13.7$) 
and best studied helium atmosphere white dwarf pulsator. A recent spectroscopic analysis by 
\citet{Bedard17} gives GD358 a spectroscopic temperature of \teff$=24,937\pm 1018$ K and \logg$7.92\pm0.05$.
However, this work relies solely on optical data.  Problems with such determinations are 
well known \citep{Bergeron11}, so we prefer to rely on the combined UV and optical temperatures
of \citet{Nitta12} and \citet{Koester2014}.  This spectroscopic 
temperature (\teff$=24000\pm500$ K) and \logg$=7.8$ places GD358 near the red edge of 
the instability strip. GD358's 
pulsation spectrum contains a series of independent radial overtones, and many have 
complex frequency structure.  For one epoch of data taken during the Whole Earth 
Telescope (WET) run XCOV25, models involving magnetic fields and oblique rotation are 
proposed to explain such structure \citep{Montgomery10}.

Since the XCOV25 WET run reported in \citet{Provencal09}, we have maintained an active 
observing program of this complex star. These new observations have successfully 
identified additional periods in GD358's pulsation spectrum, bringing the total 
known independent radial overtones to 15. Thirteen of these modes belong to a 
consecutive $\ell=1$ sequence, the longest sequence observed in a DBV. 

Paradoxically, among the DBVs with enough detected periods to be fitted 
asteroseismically, GD358 is the only one that has not been analyzed using the 
complex C/O profiles adapted and parameterized from stellar evolution 
calculations (e.g. {}\citet{Salaris97,Althaus05}). The most recent fits of GD358 
\citep{Metcalfe03b} were performed using 11 observed modes and simple models 
where the oxygen abundance drops linearly from its central value to zero. This study 
was plagued by a symmetric asteroseismic signature from the core and the 
envelopes in the models \citep{Montgomery03} and was subsequently unable to derive 
a unique fit to the period spectrum. We present here a new detailed asteroseismic 
analysis employing more sophisticated interior chemical profiles. With these profiles,
we are able to remove the degeneracy in the best fit parameters and better constrain 
the asteroseismic fits.

The present analysis also allows us to place GD358 in the context of stellar 
evolution. According to the models, the precursors of DBs emerge from the 
``born-again'' post-AGB  helium flash phase with surface envelopes composed of 
uniformly mixed helium (He), carbon (C), and oxygen (O) \citep{Dreizler98,Herwig99}. 
During the cooling process, the helium diffuses 
upward and gradually accumulates to form a chemically pure surface layer. 
This diffusion process naturally produces a double-layer envelope structure, with pure helium 
near the surface and mixed elements below \citep{Althaus09,Fontaine02, Dehner95}. 
Diffusion is not yet complete by the time the star reaches the blue edge of the DBV instability 
strip. At this point, the atmospheric structure consists of a thin helium surface layer and a deeper 
layer of mixed carbon, oxygen, and helium surrounding the carbon and oxygen core. An important testable prediction of 
this model is that for any white dwarf of a given mass, the pure helium surface layer 
will steadily grow thicker as the DB star cools. When looking at the population of 
DBVs, we would expect to find a general increase in helium layer thickness across 
the DB instability strip. GD358 is the fourth DBV we can use to check this prediction. 
The current three DBVs (\citet{Bischoff-Kim14,Sullivan08, Metcalfe03c}) paint a 
picture qualitatively consistent with the diffusion calculations, with the hotter 
best fit models having thinner pure helium layers. With GD358, we seek to 
further define this trend.

In Section \ref{data}, we present our new observations and outline the data 
reduction process. In Section \ref{freq}, we establish the framework for 
frequency identification, and present the list of frequencies used for the 
asteroseismic investigation. We perform further analysis of the observed 
frequencies in Section \ref{analysis}, and motivate the $\ell$ and $m$ 
identification of the modes. We present the asteroseismic fitting of GD358's 
pulsation spectrum in section \ref{fitting}, present the results in section 
\ref{results} and discuss our results in section \ref{discussion}.

\section{New Observations and Data Reduction}
\label{data}

GD358 was discovered in 1982 \citep{Winget82} and was a target of the 
WET in 1990, 1994, 2000, and 2006 \citep{Provencal09, 
Kepler03,Winget94}. New observations presented here include 278 individual 
observing runs (1195.1 hrs) spanning 2007-2016 (Table \ref{journal}). Each 
season of observations was obtained as part of multi-site WET campaigns 
\citep{wet90}. 

Our data reduction tracks the prescription outlined in \citet{Provencal12}. In brief, the 
new observations were obtained with CCD photometers at multiple sites.  
Each photometer has its own specific effective band pass. Nonuniform sensitivities of 
different detectors will influence the observed pulsation amplitudes in the entire data 
set. We strive to reduce these bandpass issues by using CCD photometers with similar 
detectors when feasible. If possible, we also implement a red cutoff filter (BG40 or 
S8612) in the optical path to normalize wavelength response and minimize extinction 
effects. 

We accomplished basic image reduction and aperture photometry through the Maestro 
photometry pipeline described by \citet{Dalessio2010}.  Each image is corrected 
for bias and thermal noise, and normalized by its flat field. Maestro 
automatically covers a range of aperture sizes for the target and comparison 
stars. For each individual run from each observing site, we chose the combination 
of aperture size and comparison star(s) resulting in the highest quality light curve. 

We accomplished the second phase of data reduction using the WQED pipeline 
\citep{wqed}.  WQED examines each light curve for photometric quality, removes outlying 
points, divides by suitable comparison stars, and corrects for differential extinction. 
Since we rely on relative photometry through the use of nearby comparison stars, our 
observational technique is not sensitive to oscillations with periods longer than a 
few hours. The final product from the WQED pipeline is a series of light curves with times 
in seconds and amplitude variations represented as fractional 
intensity (mmi). The unit is well known in the WET community.  It is a linear 
representation of the fractional modulation intensity (1 mmi $\approx$ 1 mmag). We 
present our Fourier transforms (FTs) in units of modulation amplitude ($1\ {\rm mma}=1 
\times 10^{-3}\ {\rm ma} = 0.1 \% = 1\ {\rm ppt}$).

\begin{figure}
\epsscale{0.7}
\plotone{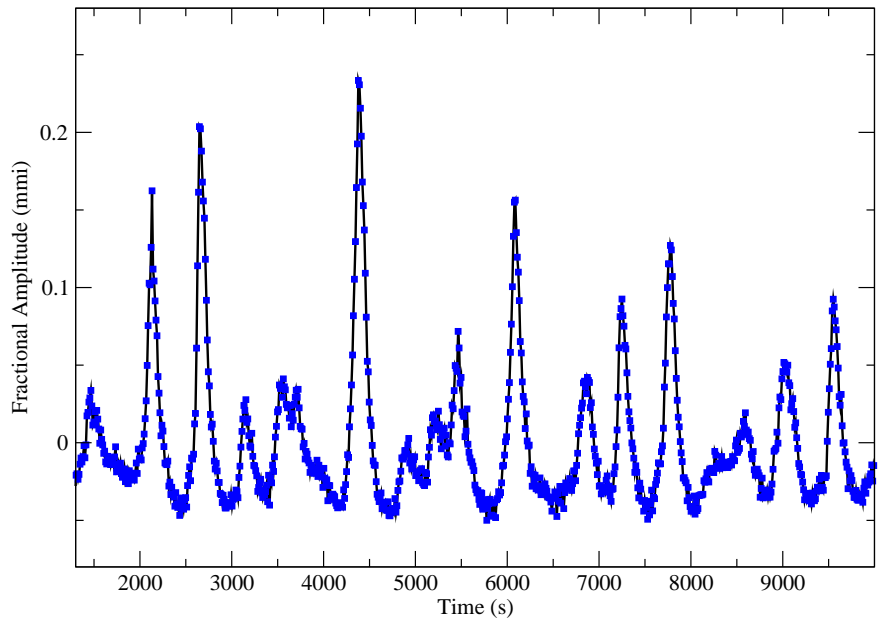}
\caption{Light curve of GD358 obtained with the Peak Terksol 2.0 m telescope.  
Each point corresponds to a 10 s exposure. The nonlinear, multiperiodic nature 
of this star is clearly evident. (A color version of this figure is available in 
the online journal.)
\label{tersk}
}
\end{figure}

Our final reduction step is to combine the individual light curves (an example 
is shown in Fig. \ref{tersk}) and apply barycentric corrections to 
create complete light curves for GD358 for each 
observing season. As we do for all white dwarf pulsators, we assume GD358 
oscillates around a mean light level. This important assumption allows us to assess 
overlapping light curves from multiple telescopes and identify and correct any 
residual vertical offsets that are instrumental in nature. The question of the treatment 
of overlapping data is discussed in detail in \citet{Provencal09}. We find no significant
differences between the noise levels of FTs  using: 1) the combination of every 
light curve including overlapping segments from different telescopes, 2) the combination 
of the subset of light curves where we retain only the highest signal to noise observations in overlapping 
segments and 3) combining all light curves incorporating data weighted by 
telescope aperture.

\begin{figure}
\plotone{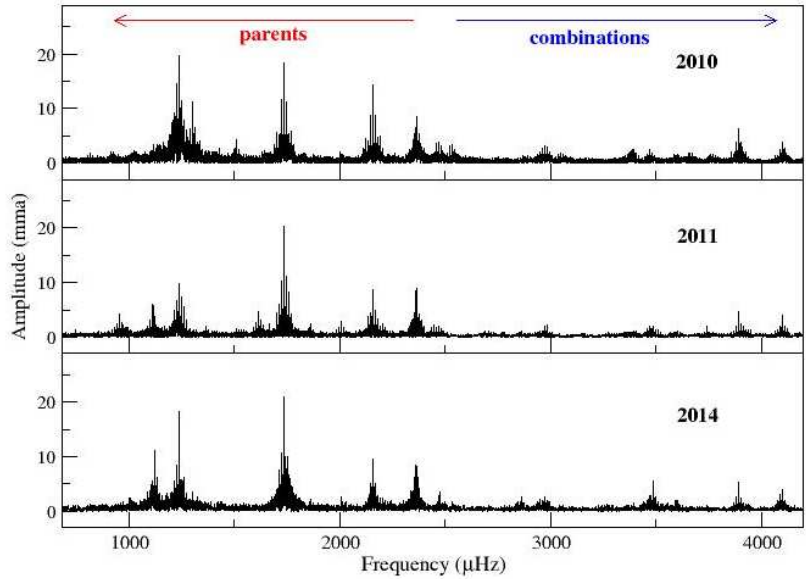}
\caption{Fourier Transforms of GD358 for the 2010, 2011, and 2014 observing 
seasons.The frequency range of the observed series of $\ell=1$ modes discussed 
in Section \ref{analysis} is indicated by the left arrow (red).  Peaks below 
$\approx$ 2400 \muHz\ are combination frequencies discussed in Section 
\ref{mode} (right arrow, blue). (A color version of this figure is available in 
the online journal.)
\label{fts}
}
\end{figure}

The complexities associated with GD358's pulsations (see Section \ref{freq}) led 
us to re-reduce all available archival data \citep{Provencal09, Bradley04, 
Kepler03, Winget94, Winget82} to insure continuity in methodology. The final 
result is a series of light curves for each observing season between 1982 and 
2016. For the new observations, 2007 contains 8.1 hrs of observation, 2008 26.1 
hrs, 2009 45 hrs, 2010 201.3 hrs, 2011 401.1 hrs, 2012 150.5 hrs, 2013 87 hrs, 
2014 184.6 hrs, 2015 55 hrs and 2016 90.6 hrs. We are limited to ground based 
facilities with inherent weather issues, so our coverage is not continuous. This 
incompleteness produces spectral leakage in the amplitude spectra. To quantify 
this, our standard procedure samples single sinusoids using the exact times as 
the original data for each season. The resulting amplitude spectrum, known as the 
``spectral window'', is the pattern produced in an FT by a single frequency sampled 
exactly as the original data. The FTs for 2010, 2011, and 2014 are given in 
Fig.~\ref{fts}.

\section{Frequency Identification}
\label{freq}
 
Our goal is to compile a complete list of GD358's observed independent and 
combination frequencies to be used in a comprehensive asteroseismic analysis. 
GD358 is well known for exhibiting changes in amplitudes and small but not 
insignificant frequency variations on a range of timescales 
\citep{Provencal09, Kepler03}.  The amplitude and frequency variations evident 
in Fig.~\ref{fts} demonstrate that it is not feasible to analyze the entire data 
set as one unit. To minimize the effects of the long term  amplitude and 
frequency variations, we analyze the light curves from each observing season 
individually.

\begin{figure}
\plotone{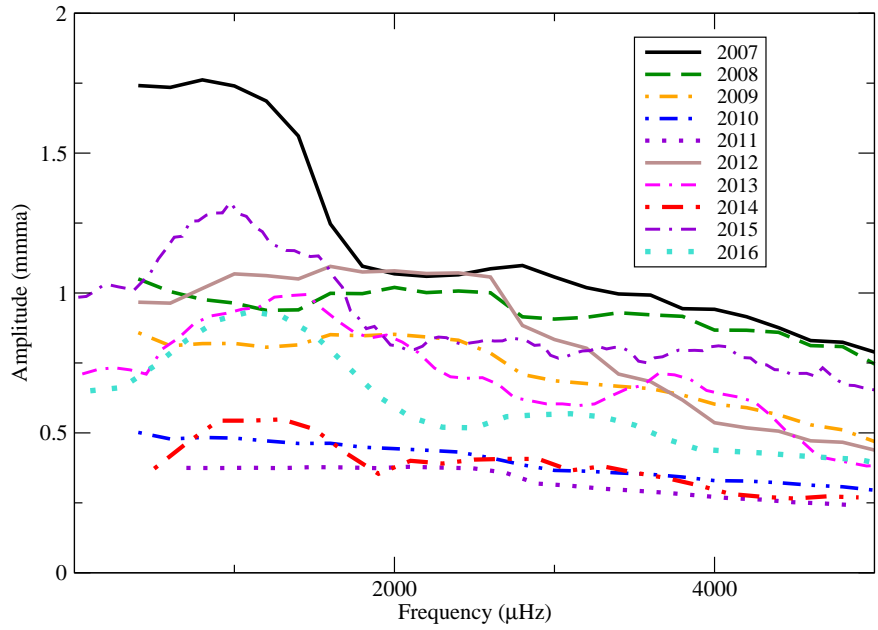}
\caption{A comparison of the average noise as a function of frequency for the 2007-2016 observing seasons.  Each data set was 
prewhitened by its dominant frequencies.  The noise levels for each season are somewhat different.  This must be taken into 
account during frequency analysis. (A color version of this figure is available in the online journal.)
\label{noise}
}
\end{figure}

We use {\sl Period04} \citep{Lenz05} for Fourier analysis and nonlinear least 
squares fitting to identify frequencies of statistical significance in each 
observing season. Our standard procedure is to adopt the criterion that any detected 
peak have an amplitude at least four times above the average noise level in the given 
frequency range \citep{Provencal12}. This criterion places a 99.9\% 
probability that the peak represents a real signal, and is not a result of 
random noise \citep{Scargle82, Provencal12}.  We define ``noise'' as the 
frequency-dependent average amplitude after removal of the dominant frequencies. 
This is unquestionably a conservative estimate, as it is impractical to assume 
that the complete set of ``real'' frequencies are removed when determining the 
noise level. This is inarguably true for GD358, where amplitude modulation is 
present, and the peaks above $\approx$ 2500 $\mu$Hz\ are combination frequencies 
(see Section \ref{mode}, \citet{Provencal09}).  Fig.~\ref{noise} displays the 
average noise as a function of frequency for the 2007-2016 observing seasons. A 
similar plot for the archival data is presented in Fig. 3 of 
\citet{Provencal09}. In addition, we performed Monte Carlo simulations using the 
routine embedded in {\sl Period04}.  This routine generates artificial light curves containing 
the original times, the fitted frequencies and amplitudes, and also includes added Gaussian 
noise. A least squares fit is performed on each artificial light curve. The estimated uncertainties
arise from the distribution of fit parameters. Our Monte Carlo results 
are consistent with our average noise estimates. 

Having established our baseline noise levels for each season, we proceed to frequency selection and identification. Our standard 
frequency selection procedure identifies the largest amplitude resolved peak in the FT, fits a sinusoid with that frequency to the data set, subtracts the fit from the light curve, recomputes the FT, examines the residuals, and repeats the process until no significant power remains. This technique is known as prewhitening. Prewhitening has inherent dangers, and must be employed with extreme vigilance. A general danger is posed by the presence of alias peaks in the spectral window.  For GD358, we are also aware of amplitude and/or frequency modulation in our data set. A detailed discussion of the prewhitening procedure and steps taken to minimize the effects of amplitude modulation is given in \citet{Provencal09}. The final frequency identifications, amplitudes, and errors for each observing season are derived from a simultaneous fit of all identified frequencies. All independent frequencies meeting our detection criteria for each observing season are given in Table \ref{freq1} and Table \ref{freq2}.

\section{Frequency Analysis}
\label{analysis}

\subsection{Frequency Distribution}
\label{freqdistr}
Perusal of Tables \ref{freq1} and \ref{freq2} shows that GD358's observed frequencies vary in two important ways: 1) frequencies detected in a given observing season are not found  in all observing seasons, and 2) the detected frequencies are not statistically identical from year to year. Asteroseismology is based on the assumption that the available pulsation frequencies are linked to stellar structure.  Since we are fairly certain that GD358's internal structure  does not change on the timescales of the each observing season, we can assume that GD358 excites different subsets of its available pulsations at different times. While this is a common phenomenon seen in white dwarf pulsators, the selection mechanism remains unknown. The best way to ascertain GD358's complete set of pulsation frequencies is to observe over multiple years and combine frequency identifications obtained from each season \citep{Kleinman98}.  

\begin{figure}
\epsscale{0.9}
\plotone{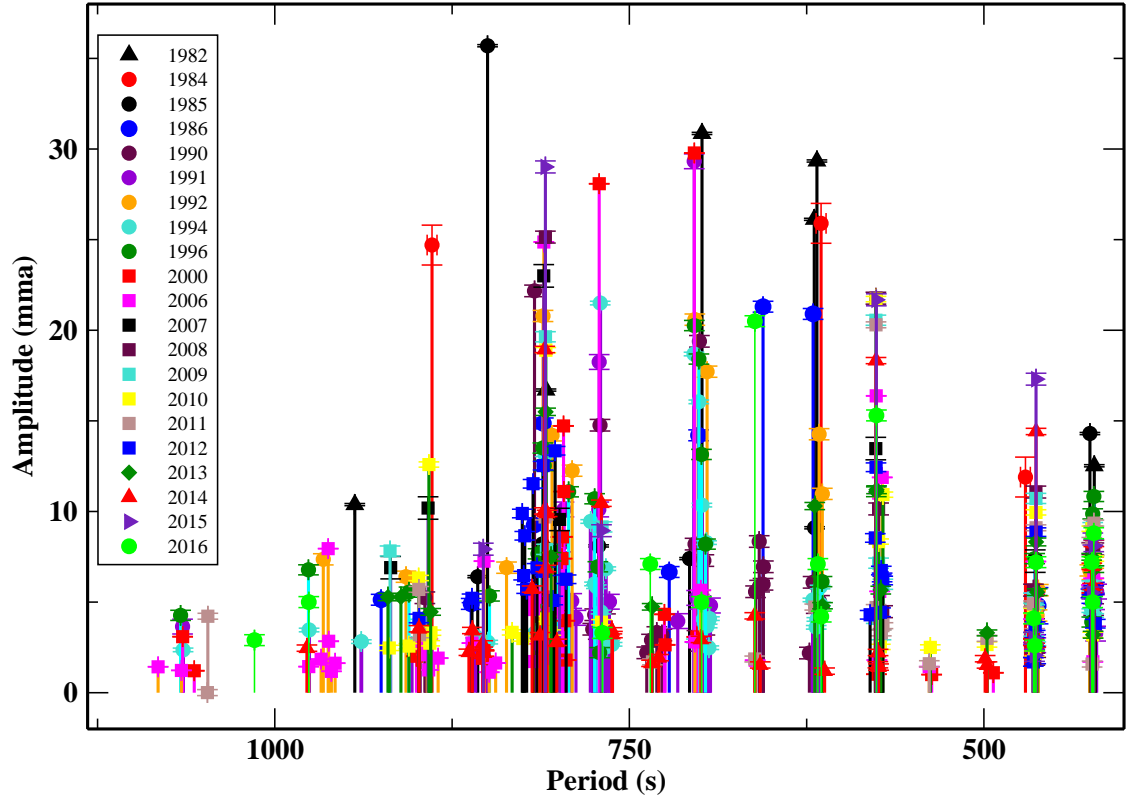}
\epsscale{0.9}
\caption{A schematic representation of GD358's pulsation modes for all available data between 1982 and 2016. Systematic patterns of distribution in amplitude and period (frequency) are evident. The bands between 1000 and 400 s (2400 and 1000 \muHz) are of particular importance for this work. Error bars (1$\sigma$) are given for amplitude and frequency.   (A color version of this figure is available in the online journal.)
\label{schematic}
}
\end{figure}

\begin{figure}
\epsscale{0.9}
\plotone{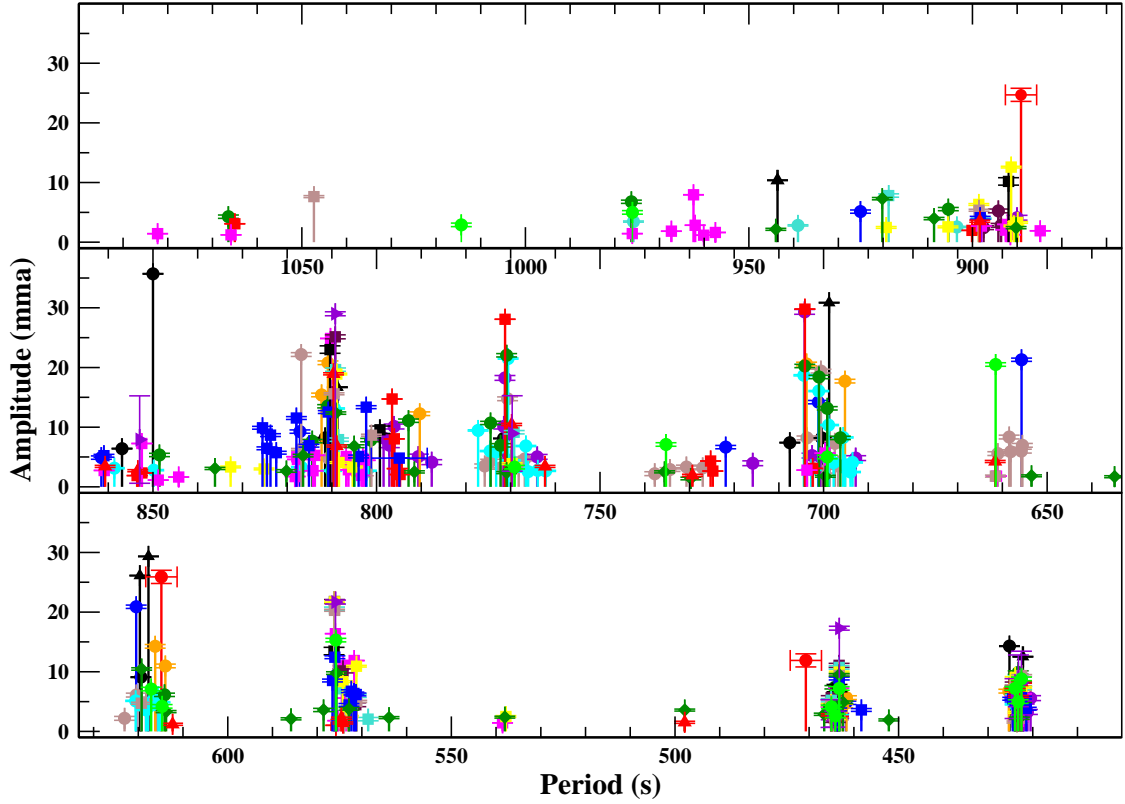}
\caption{An expanded view of GD358's frequency distribution from Fig.~\ref{schematic}}. Error bars (1$\sigma$) are given for amplitude and period (frequency). 
Period (frequency) errors are typically the size of or smaller than the plotted points. (A color version of this figure is available in the online journal.)
\label{schematicb}

\end{figure}

Figs.~\ref{schematic} and \ref{schematicb} present schematic representations of all GD358's independent frequencies detected between 1982 and 2016. We present these 
figures using period for the x-axis because g-modes in white dwarfs are roughly equally spaced in period. The figures contains all peaks in all observing seasons meeting our 
detection criteria, and so represent the observed distribution of GD358's pulsation modes. 

The features of asteroseismic importance are the localized bands between 1100 and 400 s (900 and 2400 \muHz). We interpret the bands to represent a series of modes of spherical index $l$ and radial overtone $k$. While we detect individual periods longer than 1100 s, in this region consecutive radial overtones overlap in frequency space, making detailed mode identification difficult. 

\begin{figure}
\epsscale{1.0}
\plotone{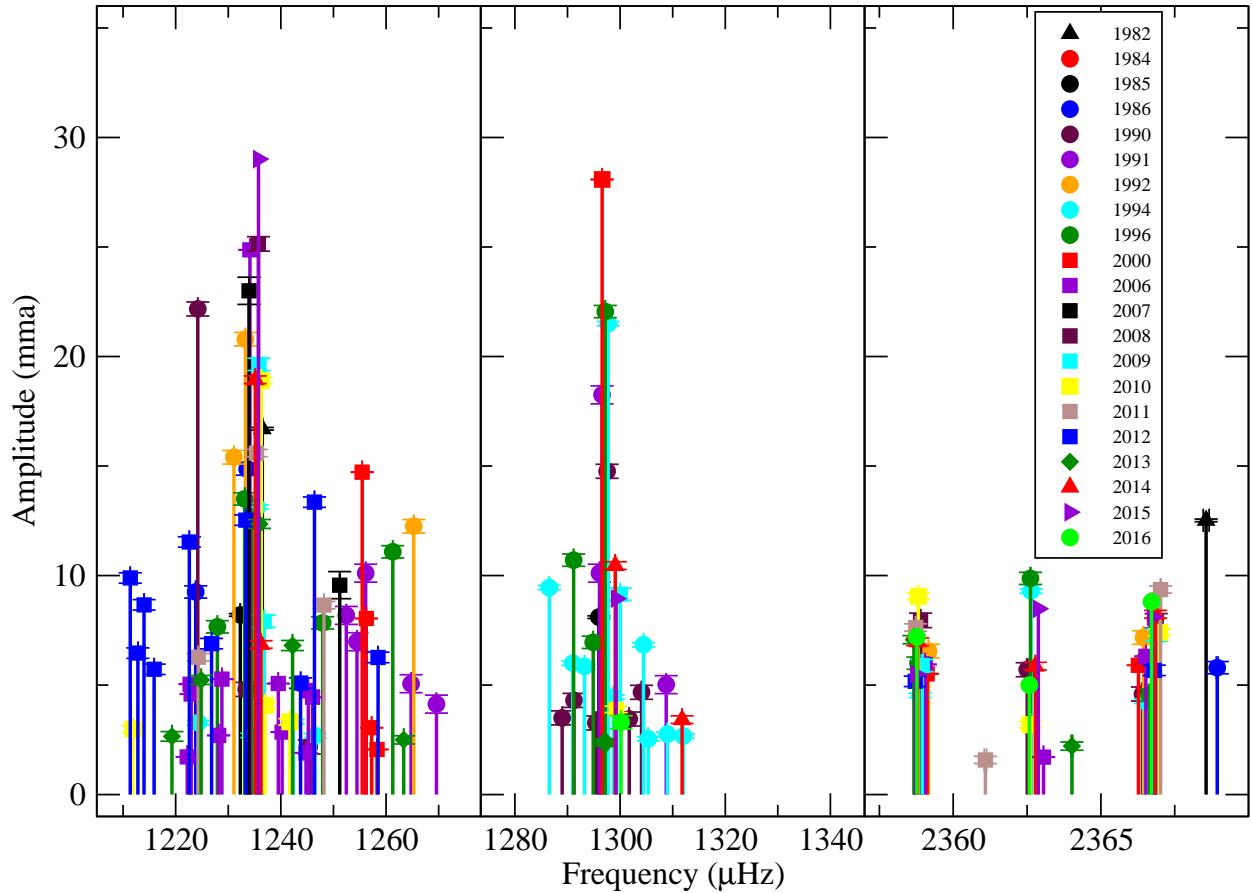}
\caption{The detailed schematic distribution of frequencies for the bands near 1233 ($k=18$) ({\em{right}}), 1300 ($k=17$) ({\em{center}}), and 2360 ($k=8$) \muHz\ ({\em{left}}). As examples of higher frequency bands, the 1233 and 1300 \muHz\ band shows no multiplet structure. In addition the 1233 \muHz\ band is twice as wide as the 1300 \muHz\ band. The 2360 \muHz\ band does show evidence of multiplet structure.   Assuming the triplet represents $\ell=1$ implies a rotation period of $\approx 1.5$ days. Note changes in y scale for all panels, and x scale for $k=8$}. Amplitude and frequency errors (1$\sigma$) are plotted for each measurement.  Please note that frequencyerrors are typically smaller than the plotted point (A color version of this figure is available in the online journal.)
\label{rot}
\end{figure}

The presence in our ensemble data set of localized bands rather than multiple detections of the same frequency was a surprising development. Each band contains frequency detections from multiple observing seasons, and is significantly wider in frequency space than the determined error of any single measurement. For example, the band near 1300 \muHz\ (770 s) (Fig.~\ref{rot}) spans 25.5 \muHz. The widest band, near 1238 \muHz\ (807 s), spans 58 \muHz. The timebase for each observing season varies from weeks to several months, resulting in individual measurement errors of much less than 1 \muHz. Further analysis of the frequencies and errors from each observing season reveals that, in most cases, the detected frequencies are stable over the timebase of each observing run. Our conclusion is that an unknown process must be acting on the frequencies, causing them to wander on timescales much longer than several months.

\begin{figure}
\epsscale{0.8}
\plotone{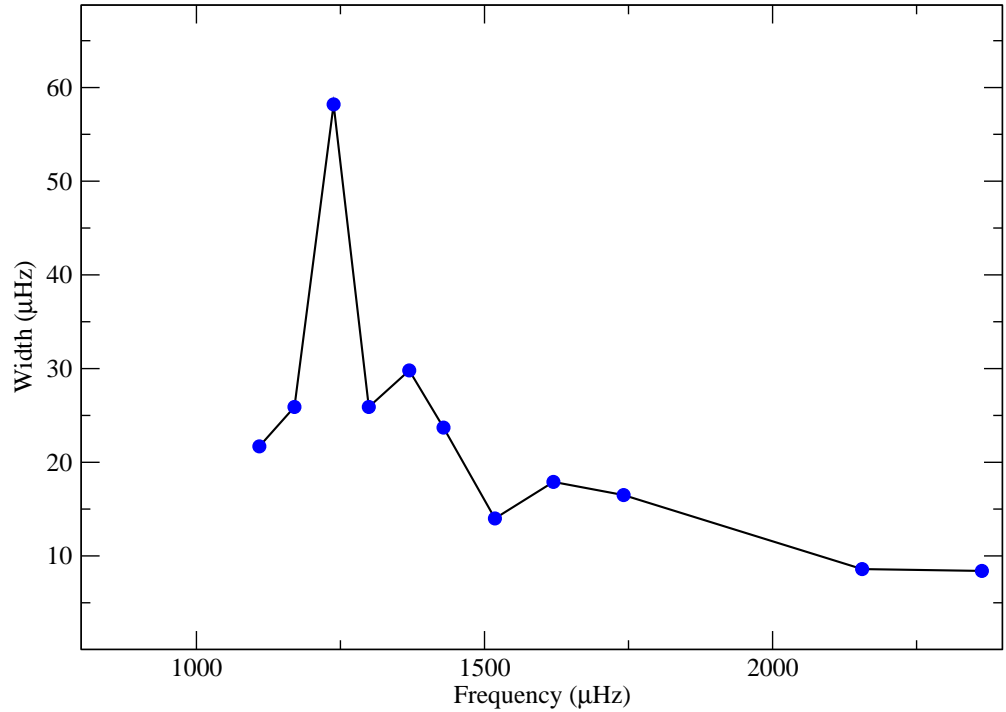}
\caption{Width in frequency space of each band of power in Fig.~\ref{schematic}. Please note that we use the widths of the triplets for k=8 and 9. We find an overall increase in band thickness with decreasing frequency (increasing period). The band at 1233 \muHz\ (811~s) has a width at least twice as large as other bands. (A color version of this figure is available in the online journal.)  
\label{thick}}
\end{figure}

The simplest explanation for these bands
that jumps to mind is rotational splitting.  However, rotational splitting predicts stable multiplet structure of a frequency width that does not change from band to band. With the exception of the highest frequency (shortest period) bands ($k=8$ and $k=9$ at 463 and 420 s from \citet{Provencal09}), we find no evidence for stable multiplets  in GD358's observed frequency distributions (Fig.~\ref{rot}). 
This does not imply that the structures reported by \cite{Winget94}, \cite{Kepler03}, \cite{Montgomery10} and others are not real. In particular, it is clear that the 1741 \muHz\ (574 s, $k=12$) mode can be explained by the oblique rotation model during 2006. However, previous work represents short snapshots of GD358. When examined over timescales of decades, these structures do not regularly recur, and so cannot be interpreted as simple solid body rotational splitting. 
We also find evidence of increasing band width as a function of increasing frequency (Fig.~\ref{thick}). The varying widths of the lower frequency bands are not the result of rotational splitting, but must be a signature of other long term processes influencing the interior propagation region as seen by each mode. In other words, an unknown physical process is modifying the internal structure of GD358 over timescales of decades.


\citet{Bell15} present an interesting analysis of similar behavior in the hydrogen 
atmosphere pulsator (DAV) KIC4552982. The authors identify 17 bands of pulsation frequencies. KIC4552982 is one of the coolest ZZ Cetis known \citep{Tremblay2013}, and 
it follows the general pattern exhibited by GD358: the highest frequency (shortest period) mode shows evidence of rotational splitting, while the lower frequency (longer period) modes are complex bands. The authors note that this DAV's bands have different widths in frequency space, with the widest band at 950 s, and that there may be astronomical significance to this. Although the observational timebases and coverage are quite different (20 continuous months for KIC4552982 vs 34 incomplete years for GD358), our data presents the opportunity to investigate this for a cool DB pulsator.  We measured the widths of each of GD358's band where we have more than 10 detections. We define ``width'' as the difference between the lowest and highest frequency detected in each band.  We find no correlation of width with number of detected peaks in each band. Interestingly, the widths of GD358's bands are commensurate with those found in KIC4552982. We find a general increase in width with decreasing frequency (increasing period), until we reach the band at 1238 \muHz\ (807 s).This band has a width at least twice as wide as any other. 


\citet{Mont2016} recently presented a scenario that could provide a natural explanation for the behavior seen in GD358 and in the coolest pulsators.
White dwarfs pulsate in nonradial g-modes. As a DBV such as GD358 pulsates, it experiences local surface temperature variations as large as 3000 K.  The temperature variation affects different modes in different ways.  An appropriate analogy is to consider each mode's propagation region as a box with a lid. The box is defined as the region where the mode frequency is less than both the buoyancy (Brunt-V{\"a}is{\"a}l{\"a}) and acoustic (Lamb) frequencies.  The box's lid represents the mode's outer turning point. For short period modes (such as $k=8$ and $k=9$ in GD358), the lid is defined by the acoustic frequency, which is relatively insensitive to surface temperature variations. As the star pulsates, the box lid is securely fastened so these modes should be stable.  For the longer period modes, the lid of the box is defined by the buoyancy frequency, which goes to zero at the base of the surface convection zone. This is the important point: for longer period modes, the box lid is actually defined by the base of the convection zone, which is very sensitive to local temperature variations.  As GD358 pulsates, its convection zone deepens and thins in response to the local temperature variations. In our analogy, the lid of the box is loose and moves, effectively changing the characteristics of the box. Long period modes with outer turning points defined by the base of the convection zone should be perturbed. For GD358, $k$=8 and $k$=9 would represent unperturbed modes with the box lid firmly fastened. The triplets here are dominated by rotational splitting. The gradual increase in mode width with frequency would represent a gradual loosening of the lid. It could be argued that the dividing line between ``thin'' (unperturbed) and ``thick'' (perturbed) modes occurs at 807 s (1238 \muHz, $k=18$). This mode is so much wider than the lower $k$ modes that it may mark the point at which the outer turning point for GD358's modes becomes ${dominated}$ by the base of the convection zone. Further investigation into this behavior in GD358 and other WD pulsators is necessary.  

An additional interesting result is seen by examining the distribution of average amplitudes for the bands (Fig~\ref{schematic}). In particular, the two bands at 538 and 498 s (1857.7 and 2007.6  \muHz) have never been observed at amplitudes above 4 mma. The band at 730 s (1369 \muHz) is also only observed at lower amplitudes. Interestingly, the band at 574 s (1741.5 \muHz) was not observed at large amplitude prior to 2006, and it has remained at high amplitude since that time. 

To summarize this section in broad brush strokes, we observe long timescale variations in the frequencies of GD358's longest period modes. 
The observed global pulsations involve the whole star, but each pulsation mode samples the star in slightly different 
ways.  Modes with lower frequencies (higher radial overtones $k$) preferentially sample the outer layers, while modes with higher frequencies (lower radial overtones $k$) 
have outer turning points that are farther from the surface, and so sample the deeper interior. It makes intuitive sense that lower frequency modes would be affected by processes confined to the outer stellar atmosphere, such as the convection zone or a surface magnetic field. We speculate that the observed band widths and pulsation amplitudes contain information about the convection zone and/or any surface magnetic field. Further investigation requires guidance from theory.  

\subsection{Mode Identification}\label{mode}
\subsubsection{Spherical Harmonic Indices}
Our current work has produced a well defined sequence of modes, adding to previous studies \citep{Montgomery10, Provencal09, Metcalfe00, Winget94}. The previous identification of these modes as a series of $l$=1 radial overtones is based mostly on the pulsation frequency distribution and limited spectroscopic analysis \citep{Kotak02, Castanheira05}. It is important to further investigate these identifications as we initiate an in depth asteroseismic investigation.  

GD358's combination frequencies provide a tool by which we can bolster $\ell$ identifications. Combination frequencies are observed in the FTs of moderate to large amplitude pulsators.  They are identified by their exact numerical relationships with parent frequencies.  The combinations themselves are not independent, but are produced via nonlinear effects associated with the surface convection zone \citep{Brickhill92, Brassard95, Wu01, Ising01}.   \citet{Wu01} lays the foundations, showing that observed amplitudes of the combination frequencies depend on geometric factors such as the $(\ell,m)$ indices of the parents and the inclination of the pulsation axis to the line of sight.  

The methods outlined in \citet{Provencal12} and \citet{Montgomery10} work best when applied to larger amplitude frequencies detected in high signal to noise data sets such as provided by extensive WET runs.  The primary reason for this is that combination frequencies are lower amplitude than their parents, and so are more difficult to detect in sparse data.  We chose the 1990, 1994, 2006, 2010, 2011, and 2014 observing seasons, and looked at pulsation frequencies with amplitudes above 10 mma.  

\begin{figure}
\epsscale{0.8}
\plotone{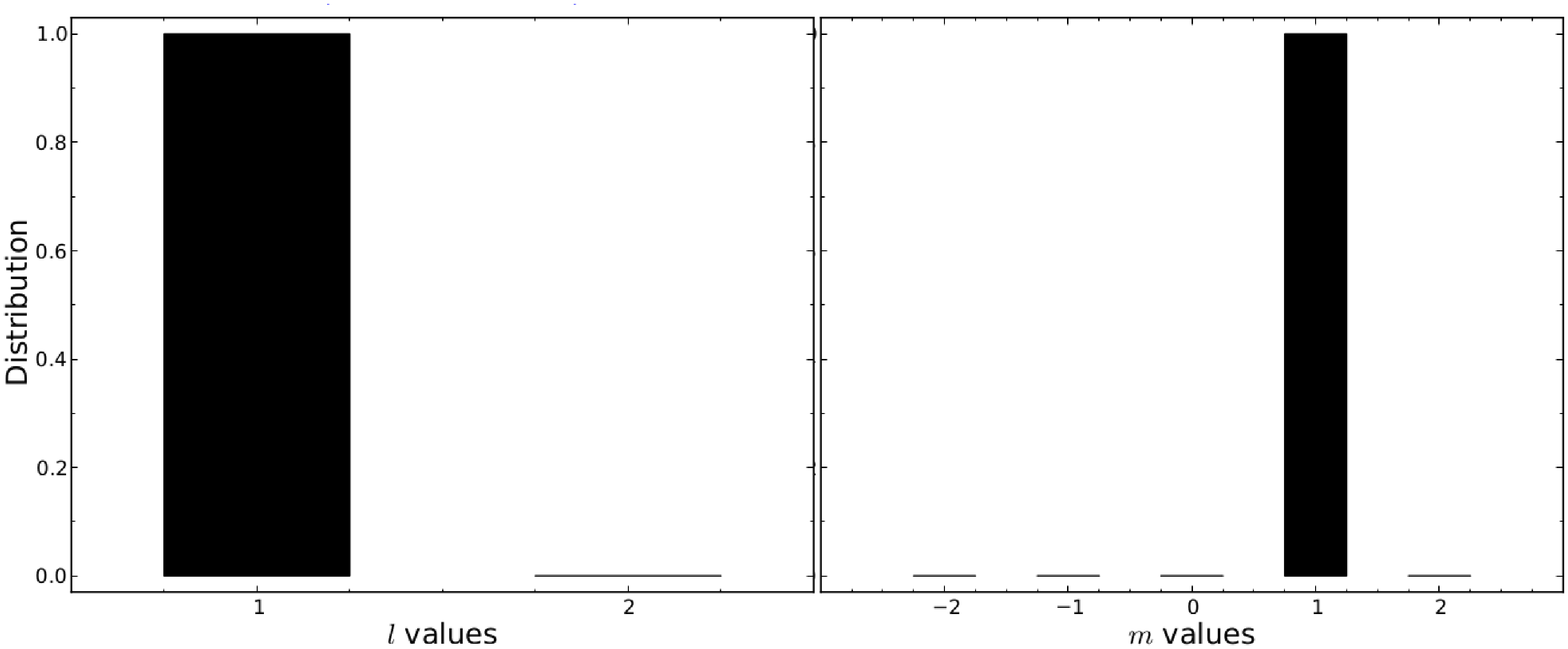}
\caption{Probability distribution (from 0 to 1) of $\ell$ and $m$ values for the 1735.96 variation detected in 2014.  The distribution is produced from 1000 simulations, selecting with $\rm{Res_{rms}}<9.5\times 10^{-6}$. The amplitudes of the observed combination frequencies argue that this is $\ell=1$, $m$=1.  (A color version of this figure is available in the online journal.)
\label{modeamps}
}
\end{figure}

As an example, Fig.~\ref{modeamps} shows the probability distribution of $\ell$ and $m$ values for the 1735.96 \muHz\ frequency as detected in 2014. To produce the distribution, we ran the amplitude code mode\_amps \citep{Montgomery10} 1000 times, and selected the results having ${\rm Res_{rms}}<9.5\times10^{-6}$, where ${\rm Res_{rms}}$ are the root-mean-squared residuals between predicted and observed amplitudes. For our example, this mode is clearly preferred as an $\ell=1$, $m=1$ mode. Table \ref{modeampstab} gives similar results for all modes above 10 mma in the 1990, 1994, 2006, 2010, and 2014 observing seasons. Combining the results from the combination frequencies with previous evidence, we are confident that the bands in Fig.~\ref{schematic} and \ref{schematicb} represent a series of $\ell=1$ modes.  

\subsubsection{Period Identifications for Asteroseismic Fitting}

\begin{figure}
\epsscale{0.8}
\plotone{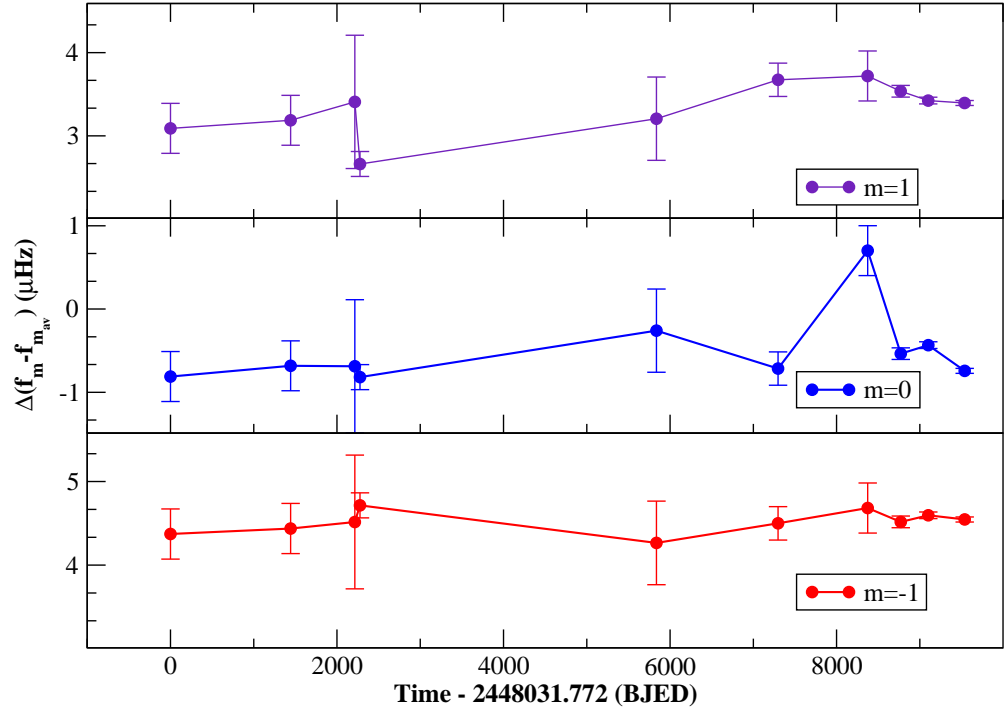}
\caption{Deviations of the multiplet components of $k$=8 from the determined average period (Table~\ref{gd}). Points represent years for which we have clear detections of each component. (A color version of this figure is available in the online journal.)
\label{k8}
}
\end{figure}

Since our asteroseismic fitting techniques require well defined periods and not extended bands, we need to determine mean periods for each band to be used in the asteroseismic fits. One way is to use the widths of the bands to determine the uncertainties by determining the FWHM. We include this determination in Table~\ref{gd}. However, the band widths may have astrophysical origins, and therefore may not be entirely due to measurement uncertainties.  It should be possible to determine mean periods with greater accuracy. First, we assume that each detected frequency in each observing season represents a separate independent determination of the underlying mode as presented at the time of observation. Secondly, we should not simply identify the largest amplitude period in each band, as amplitudes can vary on timescales of days. In Section \ref{freqdistr} we speculated that the convection zone plays a role in the intrinsic frequency variation of each band. While we do not completely understand the details of how changes at the base of the convection zone affects different modes, we have no reason to assume that this process is in any way asymmetric.  Given the astrophysical implications of the band width, the assumption of symmetry via $l$=1 identifications, and the lack of definitive multiplet structure for the lower frequency modes, it is possible to determine a central frequency for each band by averaging the detected frequencies in each band. One might question why we do not use the central components of the triplets for the shorter period ($k$=9 and 8) bands. The triplets do not have exactly equal splittings, and the central components wander in frequency over time (Fig.~\ref{k8}) \citep{Provencal09}. It is probable that some process, such as magnetic fields, is offsetting the central components. We therefore decided to treat all the bands with the same protocol. We experimented with numerous weighting techniques, and determined there is no significant difference in our solutions.  Our mean periods for each band, their uncertainties from the mean, and the FWHM for the bands are given in Table~\ref{gd}. We use these periods in our asteroseismic fitting, described below. 

\section{Asteroseismic fitting}
\label{fitting}

The basic method in our asteroseismic fitting consists of calculating grids of white dwarf models and running a fitting subroutine to match the periods of the models ($\rm P^{calc}$) with the observed periods $P^{\rm obs}$. Following standard statistical methods, each fit is assigned a fitness parameter calculated the following way:

\begin{eqnarray}
\label{fiteq1}
\sigma_{\rm RMS} = \sqrt{\frac{1}{W} \sum_{1}^{n_{\rm obs}} {w_i(P^{\rm calc}_i-P^{\rm obs}_i)^2}}, \\
W=\frac{n_{\rm obs}-1}{n_{\rm obs}}\sum_{1}^{n_{\rm obs}}w_i
\end{eqnarray}

\noindent where $n_{\rm obs}$ is the number of periods present in the pulsation spectrum and the weights $w_i$ are the inverse square of the uncertainties listed for each period in Table~\ref{gd}. We did not weigh by FWHM because not all frequency clusters have a determined FWHM. However, we tried the initial fit (see section \ref{grids}) using the subset of periods that do have a FWHM, weighing both by uncertainties and by FWHM. We found that the best fit parameters resulting from the two searches were similar. With the chosen weights, we note that the two shortest period modes have 6000 times the weight of the longest period mode. Another way to think about this is to assume that we have a calculated period that matches the highest period mode very poorly, being 20 seconds away. 20 seconds is roughly half the average period spacing for $\ell=1$ modes in the relevant area of parameter space and so it is the worse period fit one can get. In order to have the same impact on $\sigma_{\rm RMS}$, the lowest period mode would have to match to within 0.26 seconds. In essence this is almost ignoring the modes that have a period measurement uncertainty of 0.5~s or more. It is, however completely consistent with the relative uncertainty on the periods and it provides us with a true measure of the goodness of fit, while accounting for all the data we have.

\subsection{The Models}
\label{models}

To compute our models, we used the White Dwarf Evolution Code (WDEC). The WDEC uses hot polytrope models with effective temperatures above 100{,}000~K as starting points and numerically evolves them until they represent thermally relaxed solutions to the stellar structure equations and have the effective temperature of our choice. The mass and internal chemical composition profiles (i.e. no mass loss, no time dependent diffusion of elements) are fixed as an input. Each model we compute for our grids is the result of such an evolutionary sequence. The WDEC is described in detail in \citet{Lamb75} and \citet{Wood90} and the latest updates to the code as used in this work may be found in \citet{Bischoff-Kim14}. An example of an input composition profile is given in Fig. \ref{ffit1}.

\subsection{Initial Grid Search}
\label{grids}

\begin{figure}
\epsscale{1.0}
\plotone{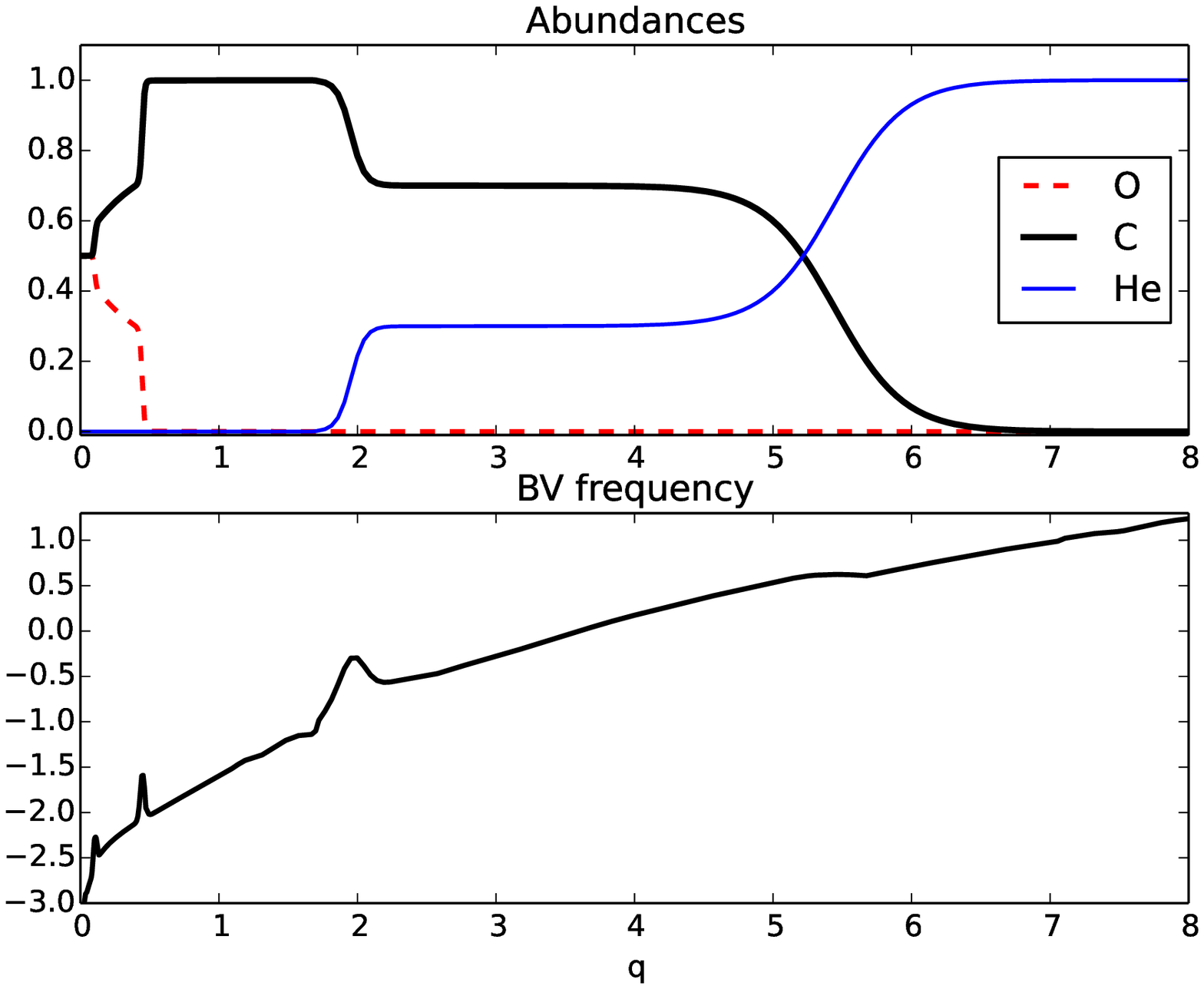}
\caption{
{\em Upper panel:} Chemical composition profiles of the best fit model ($M_{\rm env}=-2.0$, \mhe $=-5.5$, $X_{\rm fm}=0.195$, and $X_o=0.50$). The center of the model is to the left and the surface to the right. $q=2$ corresponds to $M_r = 0.99 \: M_*$. The vertical axis shows fractional abundances. 
{\em Lower panel:} The corresponding \bvf frequency ($\log{N^2}$ on the vertical axis). (A color version of this figure is available in the online journal.) 
\label{ffit1}
}
\end{figure}

In our asteroseismic fits, we initially varied six parameters: the effective temperature, the mass and four structure parameters. There are two parameters associated with the shapes of the oxygen (and carbon) core composition profiles: the central oxygen abundance (\xo) and the edge of the homogeneous carbon and oxygen core (\qfm, as a fraction of stellar mass). For the envelope structure, \menv\ marks the location of the base of the helium layer and \mhe\ the location where the helium abundance rises to $1$ (see Fig. \ref{ffit1}). \menv\ and \mhe\ are mass coordinates, defined as e.g. $M_{\rm env} = -\log(1 - M(r)/M_*)$, where $M(r)$ is the mass enclosed in radius r and $M_*$ is the stellar mass. 

As discussed in section \ref{freqdistr}, a majority of the modes observed in GD358 are affected by surface features, such as the convection zone. We expect the convective efficiency, parameterized in the mixing length parameter $\alpha$ \citep{Bohm-Vitense58} to play an important role. The role of $\alpha$ in fitting GD358's pulsation spectrum was explored and quantified in \citet{Bischoff-Kim15}. We found that in order to have an effect of $\sim$ 1 second on the goodness of fit, one had to change $\alpha$ from it's canonical value of 0.6 to 3.0 (very vigorous convection). This result is completely consistent with those of \citep{Bradley93,Bradley94a,Metcalfe02}. In the present study, we did not vary $\alpha$ but instead used 0.6.

We started with a master grid (Table \ref{fitt1}) chosen so that it covered all relevant area of parameter space and had sufficient resolution to find any region of local minimum in the fitness parameter \citep{Bischoff-Kim11a,Bischoff-Kim14}. We used the maximum resolution that was computationally manageable. The master grid involved the computation of 10,483,200 models. We fit simultaneously all 15 periods, requiring all of them to be $\ell=1$ modes. A fitness map of this initial fit is shown in Fig. \ref{ffit2}, left panel, and the parameters of the best fit model are listed in Table \ref{fitt1}.

We generated fitness maps of the sort shown in Fig. \ref{ffit2}, left panel for all of our parameters and from that, settled on the boundaries of a more restricted grid to search at a higher resolution. We show the boundaries explored in the \mstar-\teff ~plane in Fig. \ref{ffit2}, the ranges adopted for each parameter in the refined grid in Fig. \ref{ffit3}, and list those parameters in table \ref{fitt1}. We discuss how we determined the resolution to use for the refined fitting in section \ref{refinedfits}. The width of the ranges for the parameters were constrained on one end by the minimum resolution required and on the other hand by computational considerations.

\subsection{Asymptotic Period Spacing}
\label{periodspacing}

Before we refine the period-by-period fitting optimization, it is worthwhile to step back and consider what we can learn from the average period spacing of GD358. The average period spacing provides an  asteroseismic measure of the mass and temperature of the star, independent of the details of core chemical composition profiles. Higher $k$ modes are not strongly trapped in the core and according to asymptotic theory, they should be nearly evenly spaced in period. This spacing is given by \citet{Unno89}.

\begin{equation}
\label{fiteq2}
\Delta P = \frac{\pi}{\sqrt{\ell(\ell + 1)}}\left[\int_{r_1}^{r_2}\frac{N}{r}dr\right]^{-1},
\end{equation}

\noindent
where $r_1$ and $r_2$ are turning points of the mode and N is the Brunt-V\"{a}is\"{a}l\"{a} frequency. The asymptotic period spacing is $\ell$ dependent, with higher $\ell$ modes having smaller spacing. In the case of GD358, we have a single $\ell=1$ sequence so we only need to worry about the dependence of the asymptotic period spacing on the \bvf frequency. Much if not all of GD358's pulsation spectrum is close to the asymptotic limit, because the shortest period observed is a $k=8$ mode.

The dependence of $\Delta P$ on the \bvf frequency leads to higher mass and lower temperature models having a smaller period spacing (their interior is less compressible). This effect appears in asteroseismic fitting of white dwarfs and also sdB stars (also g-mode pulsators) as a ubiquitous diagonal trend in contour maps of the quality of the fits in the mass-effective temperature plane \citep[e.g][]{Bischoff-Kim14,Castanheira09,Charpinet08,Giammichele16}. One requirement for the periods of the model to match the observed periods is that the average period spacing in the models match the average period spacing in the observed pulsation spectrum. If a good match occurs for a given mass and effective temperature, then models with lower mass but higher effective temperature will also match well.

We use the sequence of 13 consecutive $\ell=1$ modes found in GD358's pulsation spectrum to calibrate our models (Table~\ref{gd}). Using these modes, we compute an average period spacing of 39.9 seconds. We call this $\Delta P_{\rm obs}$. For each model in the master grid, we compute an asymptotic period spacing ($\Delta P_{\rm calc}$). In both cases (observed spectrum and models), this asymptotic period spacing is calculated by first discarding the 10 lowest $k$ modes. 

The exact value of 10 is somewhat arbitrary, but it is chosen so that the modes we use in our computation are indeed in the asymptotic limit. The higher $k$ modes show weaker trapping than the lower $k$ modes. We then fit a line through the set ($k_i$,$P_i$). The slope gives us the asymptotic period spacing. For the models, we also calculate the residuals of the fits and discard the models that have residuals above a certain limit. The limit is chosen by checking the procedure by eye on a few models. 

\begin{figure}
\epsscale{1.0}
\plotone{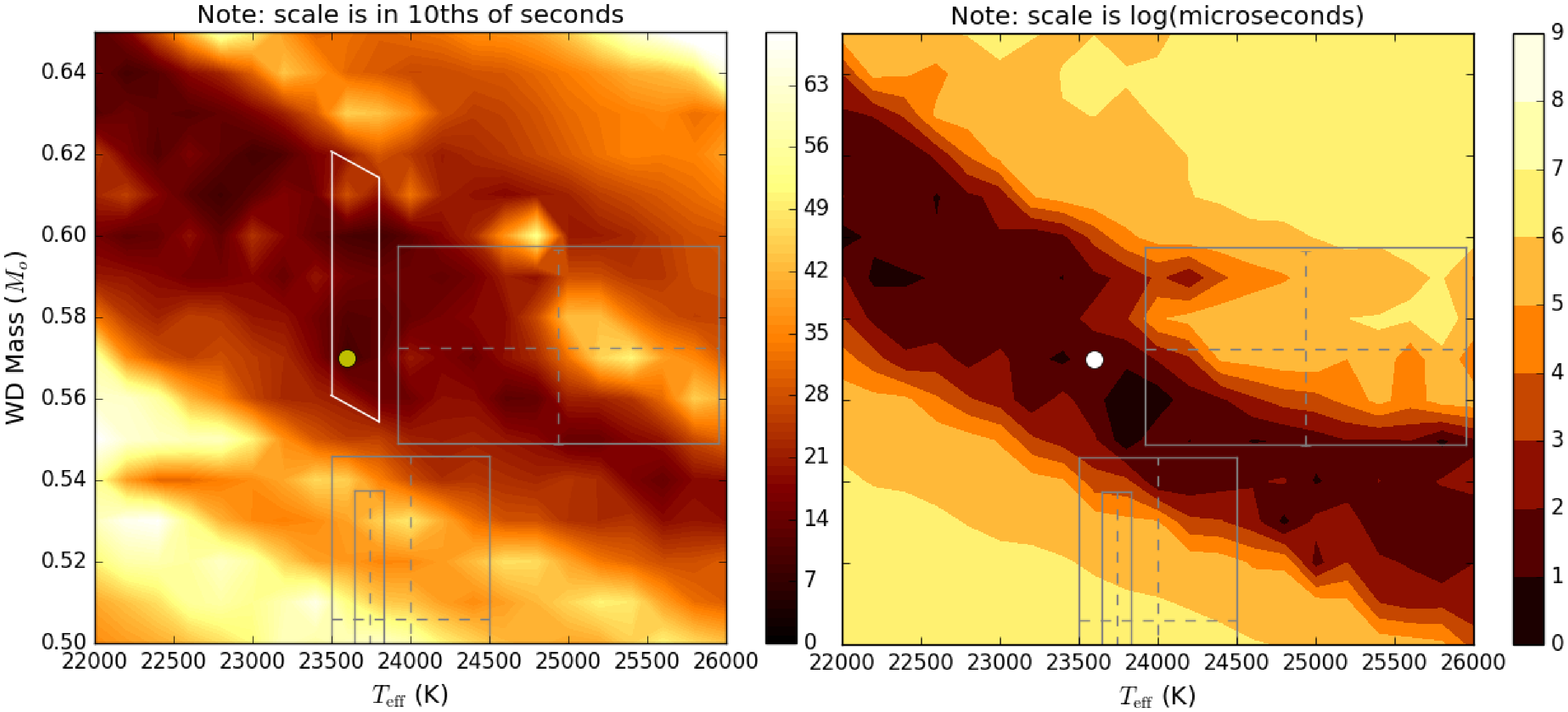}
\caption{
{\it Left panel}: Contour map showing the location of the best fit models. The quantity plotted in the mass-effective temperature plane is the fitness parameter defined in equation \ref{fiteq1}. 
{\it Right panel}: Contour map of the difference between the average periods spacing determined from GD 358's periods and the asymptotic period spacing of the models, in the mass-effective temperature plane. The worse matches pictured on the plot have $\left|\Delta P_{\rm obs}- \Delta P_{\rm calc}\right| \sim 5 \; \rm s$. We plotted as boxes the spectroscopic mass and temperature determinations of \citet{Bedard17} based on optical spectra (higher mass), and that of \citet{Koester2013} and \citet{Nitta12} based on UV spectra (lower mass). The dot corresponds to the best initial fit model. The diagonal box shows the boundaries of the refined grid (see section \ref{refinedfits}) (A color version of this figure is available in the online journal.) \label{ffit2}
}
\end{figure}

We show a contour map of the location of the models that best match the average period spacing of 39.9 seconds in Fig. \ref{ffit2}, right panel. We place it side by side with a contour map showing the location of the best fit model in the same region of parameter space, based on the master grid fitting described in section \ref{grids}. Note how the best fit model falls right within the valley where the period spacings between GD358 and the models match. This should come as no surprise, as in order for 15 periods to fit reasonably well, the model period list should have a spacing similar to that of the observed period spectrum. The recent spectroscopic determinations of \citet{Koester2013} and \citet{Nitta12} are far off the valley, at 24,000~K and $0.506^{+0.034}_{-0.051}$ \msun. Other spectroscopic determinations, such as that of \citet{Bedard17} and \cite{Bergeron11}, fall within the swath where the period spacings match.

One can fit simultaneously the average period spacing and the individual periods formally while performing the fits by using some prescription to calculate the goodness of fit. This leads to a more complex relation than defined in equation \ref{fiteq1}. Note that the period spacing is a much weaker constraint than the individual period fit. If one takes 5.0 seconds as an upper limit for goodness of fit, that includes 4\% of the models in the period-by-period fit plot (left panel in Fig \ref{ffit2}), but the entire parameter plane for the average period spacing fit plot (right panel). 

This essentially limits our search to models that already match the average period spacing. We avoid sophisticated schemes to calculate the fitness parameter and simply use equation \ref{fiteq1}.

\begin{figure}
\epsscale{1.0}
\plotone{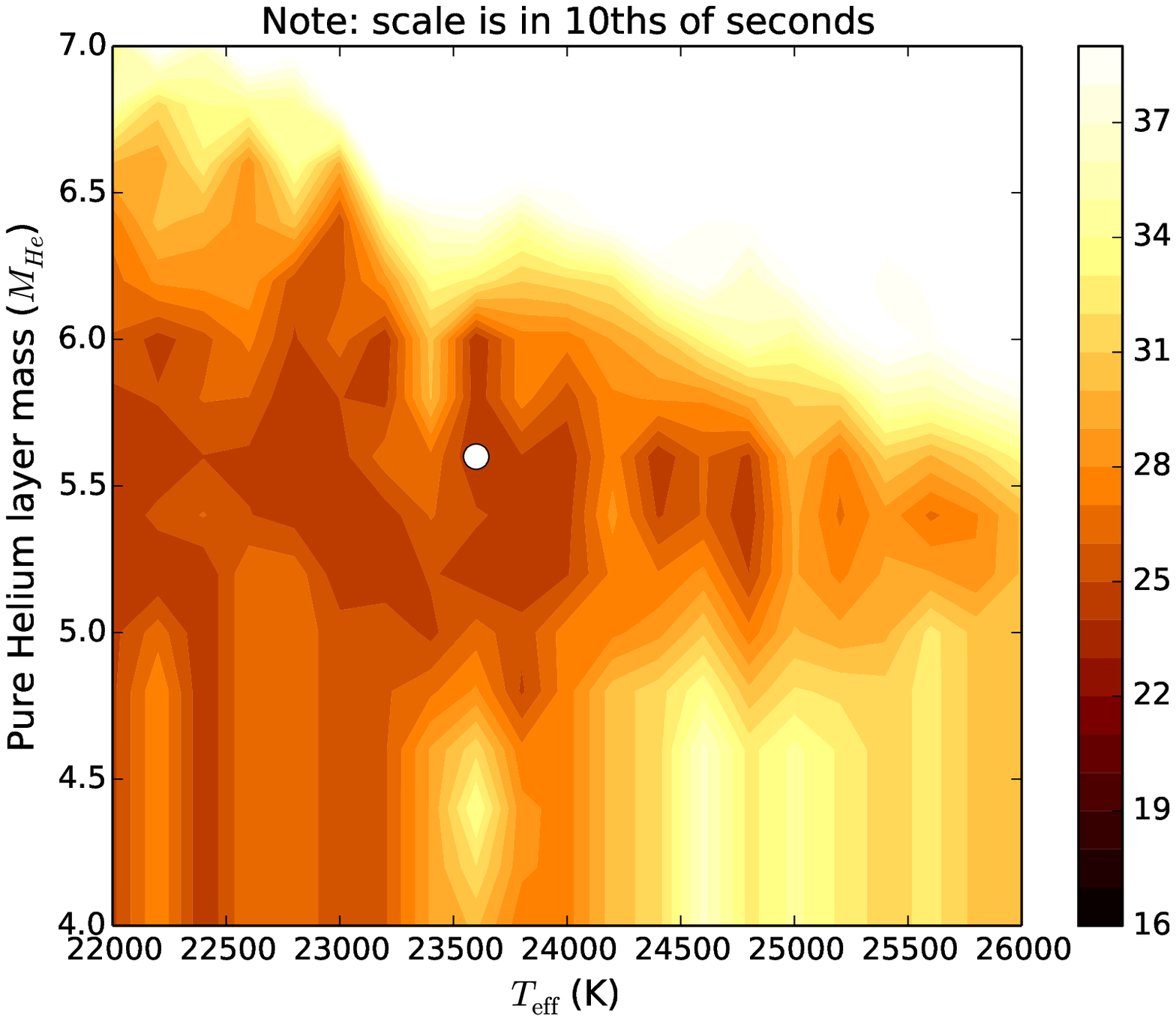}
\caption{
Contour map showing the location of the best fit models in the pure helium layer thickness-effective temperature plane. The "corners" in the contours arise from our grid resolution of 50~K and 0.1~dex. (A color version of this figure is available in the online journal.) Contour map showing the location of the best fit models in the pure helium layer thickness-effective temperature plane. (A color version of this figure is available in the online journal.) 
\label{ffit6}
}
\end{figure}

One key interior parameter we are trying to measure is the thickness of the the pure helium layer. From our initial fitting we find that the value of this parameter is relatively insensitive to the best fit value of the other parameters, as illustrated by the nearly horizontal trend between \mhe ~of 5 and 6 in Fig. \ref{ffit6}.

\subsection{Optimal Grid Resolution}
\label{refinedfits}

Having determined a more restricted region of parameter space to search for the best fit models, we now turn to the question of how fine we need to make our refined grid. We want to have a high enough resolution grid that we can be sure we capture a true minimum, but on the other hand, there are computational limitations to how many models we can afford to calculate, save, and process.

\begin{figure}
\epsscale{1.0}
\plotone{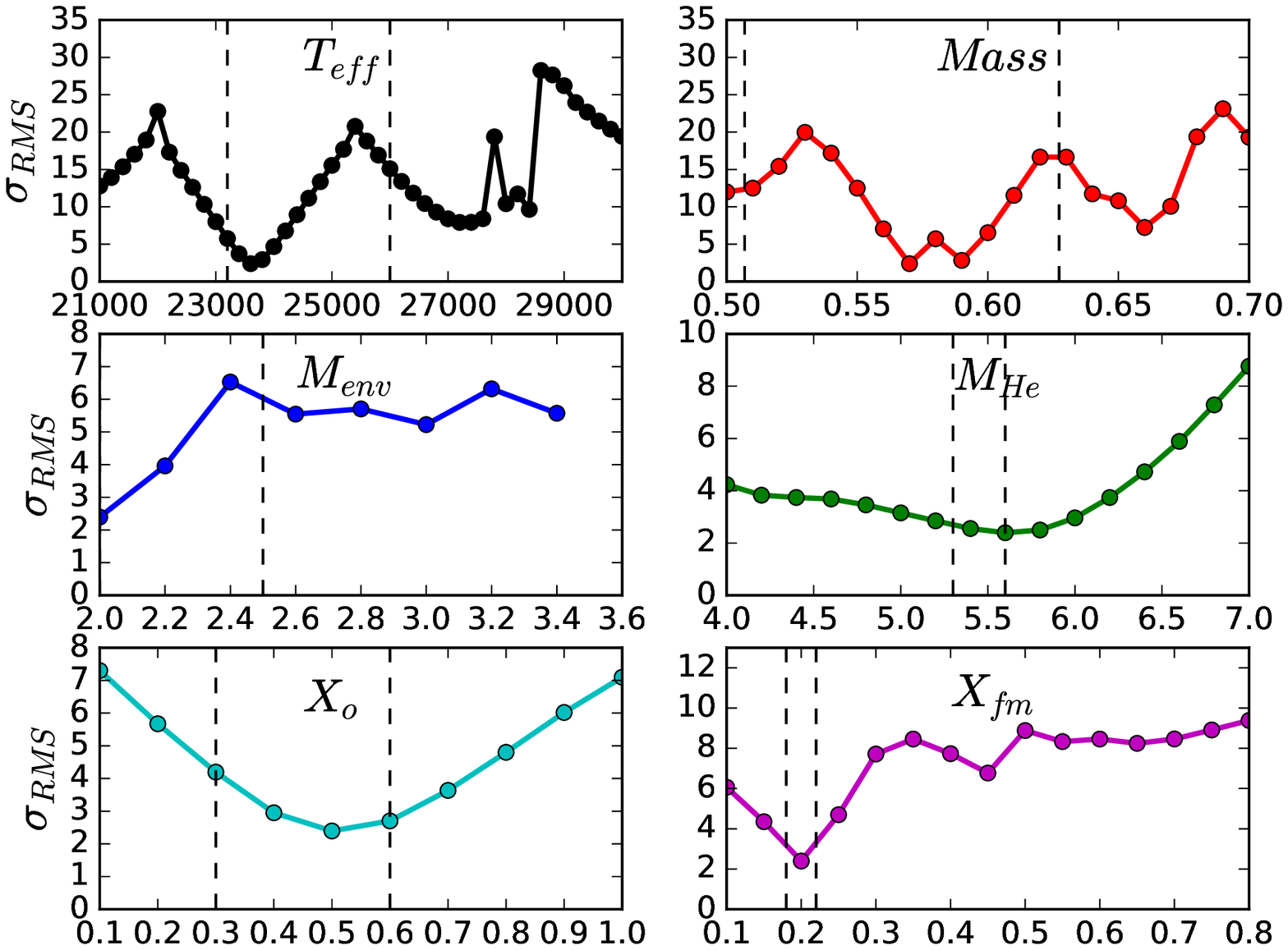}
\caption{
Dependence of the fitness parameter on different parameters. The 5 parameters other than the one shown on the horizontal axis in each plot are held to fixed values corresponding to the initial best fit model (table \ref{ffit1}). The vertical dashed lines mark the ranges considered in the refined fitting. We note that the boundaries of the refined grid are not square in the \mstar-\teff plane (see Fig. \ref{ffit1}). For the envelope mass, we went from \menv = -2.0 to -2.5. Decisions regarding the ranges to use in the refined fit were made by examining contour maps. (A color version of this figure is available in the online journal.)
 \label{ffit3}
}
\end{figure}

One way to gain a sense of how fine the grid needs to be is to make single parameter cuts through parameter space. Fig.  \ref{ffit3} shows such cuts for master grid models. The plot was made by fixing 5 of the 6 parameters to the best fit values of the best initial fit (see table \ref{fitt1}). For some parameters, the fits seem to settle to a minimum in a smooth way, while for others, they exhibit jumps. For instance, the spike in the effective temperature plot at 27,800~K is due to a period (around 530~s) that goes away and then comes back. The model with the missing period fits poorly compared to the models on either side of it. The jump from \sigrms $\sim$ 5 s to \sigrms $\sim$ 30 s that happens from 28,400~K to 28,600~K is due to a discontinuous change in the period spectrum of the models. In this case, the appearance of a new mode between the 710 and 784~s modes of the 28,400~K model. 

Discontinuities like these are unsettling, as it is easy to see that one might miss a best fit model if the grid is not fine enough. In order to quantify "fine enough", we ran systematic scans, computing models with 5 parameter fixed and allowing the 6th parameter to vary in very fine steps. We went down in step sizes to $\Delta {\rm T_{eff}}=1$~K, $\Delta M_*=0.0001 M_\odot$, $\Delta M_{\rm env}= \Delta M_{\rm He}=0.01$ dex, $\Delta X_o=0.01$, and $\Delta X_{\rm fm}=0.001$. Regardless of the behavior of individual modes in the models as the parameters vary, we can assert what step sizes will allow us to minimize our risk of missing a minimum, while minimizing the number of models to compute. We settled on step sizes of $\Delta {\rm T_{eff}}=50$~K, $\Delta M_*=0.005 M_\odot$, $\Delta M_{\rm env}=\Delta M_{\rm He}=0.1$ dex,  $\Delta X_o=0.1$, and $\Delta X_{\rm fm}=0.005$ for our refined grid.
		
\section{Results of the Period Fitting}
\label{results}

The parameters for our best fit model based on the refined grid are listed in Table \ref{fitt1} and the periods of that model in Table \ref{gd}. We find ${\rm T_{eff}}=25,630$~K, $M_*/M_\odot=0.571$, \menv $=-2.0$, $M_{\rm He}=-5.5$, $X_o=0.50$, and $X_{\rm fm}=0.195$. The goodness of fit of the model is \sigrms$=0.9361$~s. We also list the Bayes Information Criterion (BIC) number, a statistic that normalizes the quality of fits by number of free parameters and number of constraints for comparison with other studies. For a discussion applied to this parameter study, see \citet{Bischoff-Kim11a}.  One should keep in mind that the quality of fit is affected by the fact that some of the observed periods have large uncertainties (see Eq. \ref{fiteq1}).

We show the interior structure of the best fit model in Fig. \ref{ffit1}, with the corresponding \bvf frequency profile. We find that among the internal structure parameters, the thickness of the pure helium layer is well constrained, as evidenced by the horizontal trend in the contour plot in the \mhe -\teff ~ plane (Fig. \ref{ffit6}). Our result of $M_{\rm He}=5.5\pm 0.5$ is consistent with \citep{Bradley94a,Dehner95,Fontaine02} who all obtain $M_{\rm He}$ between $5.8$ and $5.5$. We also see firm evidence of a transition zone where the helium fraction drops to zero ({\menv}), which in our case is located at $2.0\pm 0.1$. This is significantly deeper than  \citep{Dehner95,Fontaine02}, but neither paper considered the sensitivity to core composition. \cite{Dehner95} used a notional $X_o =0.5$ core, while \cite{Fontaine02} used pure carbon cores. \cite{Metcalfe00,Metcalfe01,Metcalfe03} found solutions where $M_{\rm He}=2.0$ to $2.74$, which is consistent with our location of the deeper helium transition region.

We are also able to compare our core composition results with those of other authors, along with considerations from stellar evolution models and recent determinations of the $^{12}{\rm C}({\alpha},{\gamma})^{16}{\rm O}$ reaction rate (Table \ref{comparefits}). \cite{Bradley94a,Metcalfe00,Metcalfe01} all obtain $X_o=0.80$ to $0.84$, while \cite{Metcalfe03} obtain $X_o =0.67$. All of these are significantly higher than our result of $X_o=0.50$. Besides this study, only \cite{Metcalfe02, Metcalfe03} consider core C/O profiles that are scaled versions of those from stellar evolution theory, but these were based on the frequencies that were known at the time, and they typically used the NACRE \citep{Angulo99} $^{12}{\rm C}({\alpha},{\gamma})^{16}{\rm O}$ reaction rate. We also compared our oxygen abundance with the more recent results of \cite{DeGeronimo17} taking into consideration the more recent $^{12}{\rm C}({\alpha}, {\gamma})^{16}{\rm O}$ reaction rate results of \cite{deBoer17}. If we consider Fig.~7 of \cite{DeGeronimo17} and take into account the cross section uncertainties of \cite{deBoer17}, we estimate that $X_o \sim 0.5$ to $0.7$ and $X_{\rm fm} \sim 0.6$. Our oxygen fraction is on the low side of the \cite{DeGeronimo17} fraction, and would imply that the $^{12}{\rm C}({\alpha},{\gamma})^{16}{\rm O}$ reaction rate lies towards the low side of the uncertainty range. However, our $X_{\rm fm} =0.195$ is much smaller than the $0.5$ to $0.6$ of other studies, such as \cite{Giammichele18}. Resolving this discrepancy will be the subject of further studies. 

It is worth emphasizing that our model fit for GD358 is the first to consider all of the following: core oxygen fraction, location of the core transition region, the inner and outer helium transition regions, stellar mass and effective temperature. We also use more modes (15) than anyone else, along with period uncertainties that reflect changes in the pulsations in GD~358. Other studies use fewer modes and the periods are typically from one observing run, where the time-dependent changes in the periods (and their effect on the uncertainties) are not included.

\subsection{Validation of the Fitting Method and Error Estimation}

With the fit variations we discovered in section \ref{refinedfits}, it is only natural to 
be concerned about whether we have truly found a best fit. We performed a simple 
test to validate our fitting method, which consisted of using the exact same 
procedure to find a best fit to the periods of a model that was not on any of 
the grids we calculated, but that did have parameters that were very close to 
the best fit model. We used the period list for a model with parameters ${\rm 
T_{eff}}=25,630$~K, $M_*/M_\odot=0.57065$, \menv $=-2.05$, $M_{\rm He}=-5.55$, 
$X_o=0.52$, and $X_{\rm fm}=0.192$, including only the subset of 15 $\ell=1$ 
periods that match GD358's pulsation spectrum.

We first performed a fitting of the periods using the master (coarse) grid 
described in section \ref{grids}. This placed the best fit model in the 
appropriate region of parameter space. Then we refined our fits, using the grids 
described at the end of section \ref{periodspacing}. We are able to recover the 
best fit parameters adequately, with all top 5 best fit models having parameters 
${\rm T_{eff}}=25,000$~K, $M_{\rm env}=-2.1$, \mhe $=-5.5$, $X_o=0.50$, and 
$X_{\rm fm}=0.190$ and a mass ranging between  0.5730 and 0.5734 \msun. The best 
fit model has \sigrms $=0.31$~s. We remind the reader that the step sizes in the 
second phase of the fitting are 50~K for ${\rm T_{eff}}$, 0.0001 \msun ~ for 
stellar mass, 0.1 dex for helium layer masses, 0.1 for $X_o$, and 0.005 for 
$X_{\rm fm}$.

This test also allows us to place minimum error bars on at least some of the 
parameters found: $\pm 600$ ~ K in effective temperature, 0.05 dex on 
\menv\, and \mhe, 0.02 on $X_o$, and 0.002 on $X_{\rm fm}$.

We also estimated the error on the best fit parameters due to the fact that the periods to fit have uncertainties associated with them. A formal error analysis using Monte Carlo simulations based on the list of models used in the scanning tests of section \ref{refinedfits} yields ${\Delta T_{\rm eff}} = 292$ K, $\Delta M_*/M_\odot=0.006$, $\Delta M_{\rm env}=0.24$ dex, $\Delta$ \mhe~$= 0.67$ dex, $\Delta X_o=0.23$~dex, and $\Delta X_{\rm fm}=0.104$. In the simulations, we generated sets of periods based on the means and widths quoted in table \ref{gd}. We only used the 11 periods for which we were able to determine a HWHM. We did 1000 trials for each simulation and found that to be sufficient to give us normally distributed results from which we could determine standard deviations for our error estimates.

\section{Discussion and Conclusions}
\label{discussion}

We analyzed archival data and over a thousand hours of new observations for GD358, together covering a span of 34 years. With data spanning such a long period of time, we learn about the stability of the different modes. We find that the shorter period modes tend to be more stable, while the longer period modes tend to vary more in frequency over time. \citet{Bell15} have observed and modeled such stochastic behavior in KIC 4552982, a red edge DAV observed nearly continuously for 1.5 years with \emph{Kepler}. In that star, the 361.58 s triplet has sharply defined peaks, while the rest of the modes, with much higher overtone numbers, are less stable. The stability of the low k modes is likely due to the fact that they are more strongly trapped in the core. Unlike their higher k counterparts, they are affected very little by surface effects, such as the convection zone.

In analyzing the data, we found 4 new modes, adding to the 11 modes known previously. With these 15 modes, we performed a new asteroseismic fit of GD358 with models that include carbon and oxygen core composition profiles based on the stellar evolution models of \citet{Salaris97}. We find a best fit effective temperature of $23,650\pm600$~K and a mass of $0.5706\pm0.001$ \msun. While the temperature is close to the recent spectroscopic determination of 24,000~K by \citep{Koester2013}, the mass is more than one sigma above the spectroscopic mass. On the other hand, the mass matches almost exactly that found by \citet{Bedard17} and \citet{Bergeron11} (but our effective temperature is $1.3\sigma$ below their value). The spectroscopic data point of \citet{Koester2014} and \citet{Nitta12} is somewhat off the asymptotic period spacing trend (Fig. \ref{ffit2}, right panel).

With parallaxes now available from Gaia's DR2 \citep{Gaia18}, it is also useful to translate our results into a distance. For our best fit model (${\rm T_{eff}}=23,650$~K, ${\rm log(L/L_\odot) = -1.2874}$,  apparent visual magnitude = 13.65 bolometric correction =  2.45), we find a distance of 44.5 pc. For a broad range of models (21,000 to 30,000~K), the minimum possible distance is 41.5 pc, while the largest distance is 52.5 pc. The range ecompasses the distance according to Gaia's parralax ($43.101 \pm 0.055$ pc). The values are also consistent with the work of \citet{Bradley94a} and others, except that the inferred distance given by \cite{Bedard17} is slightly more than $1 \sigma$ smaller.

The presence of stochastic behavior in GD358 adds complexity to our analysis, in particular in the determination of the frequencies for the high \textit{k} modes/bands. In this work, we have chosen the simplest approach by assuming a symmetric process is responsible for the bands and calculating average values.


Numerical experiments of the type presented in section \ref{refinedfits} have shown that, consistent with the theory of non-radial oscillations, the shapes of the transition zones matter as much as where they occur \citep{Bischoff-Kim15}. Modern asteroseismic fitting vary parameters attached to the shape of these transition zones \citep{Giammichele18}. We refrained from doing this in the present study because we wanted to compare our results with previous studies of DBVs. It is unclear how much of an effect on structure parameters a change in parameterization would have.

In fitting GD358, we discovered sudden changes in the value of the goodness of fit for small changes in a given fitting parameter. \citet{Bischoff-Kim17} have explored this further by doing a more thorough period-by-period study and found that this was a manifestation of avoided crossings noted elsewhere in the literature \citep[e.g.][]{Bradley91}.

\begin{figure}
\epsscale{0.7}
\plotone{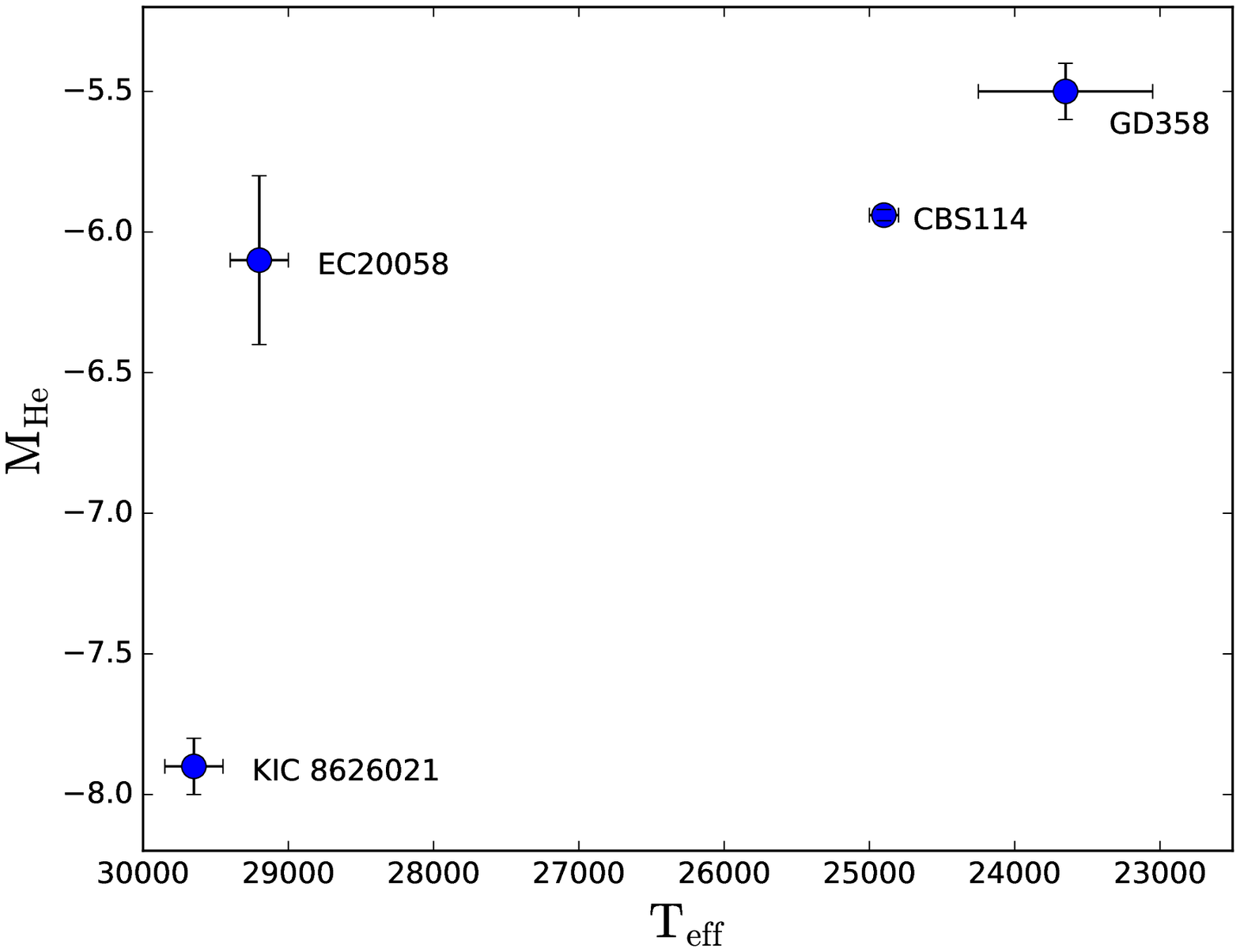}
\caption{
Relationship between effective temperature and pure helium layer mass in 4 DBVs fitted using similar models as GD358 in the present study. (A color version of this figure is available in the online journal.) \label{ffit5}
}
\end{figure}

In this study, we varied the parameters that have traditionally been varied in this type of asteroseismic fitting, so that we can place our results side by side with other studies of DBVs, including KIC 8626021 \citep{Bischoff-Kim14,Giammichele18}, EC20058 \citep{Bischoff-Kim11a}, and CBS114 \citep{Metcalfe05a}. We now have 4 DBVs that were the object of asteroseismic fitting that used a consistent set of models. We find that cooler best fit models have thicker pure helium envelopes (Fig.~\ref{ffit5}), in accordance with the outward diffusion of helium over time. To be consistent, we used the effective temperature found through the asteroseismic fitting of each star. Stellar mass also comes into play in diffusion. However, the models range in mass from 0.525 (EC20058) to 0.640 \msun ~ (CBS114), while \citet{Althaus09} show little variation in the thickness of the pure helium layer between a 0.515 \msun ~ and a 0.870 \msun ~ model. More work would be required in order to assess the significance of the trend observed with these 4 DBVs. We note that while \citet{Giammichele18} confirm the temperature determination of \citet{Bischoff-Kim14} for KIC 8626021, they find a somewhat larger helium layer mass (at -6.40), which remains thinner than the canonical value. 

\acknowledgments

We would like to thank Detlev Koester for providing bolometric corrections and other atmospheric model parameters. We would also like to thank members of the WET collaboration not on the author list who contributed helpful feedback, in particular S.O. Kepler. M.H.M. acknowledges support from NSF grant AST-1312983. Zs.B. acknowledges the support of the Hungarian NKFIH Grants K-115709 and PD-123910.

{\it Facilities:}  \facility{Struve()},
\facility{KPNO:2.1m ()}, 
\facility{}, \facility{BOAO:1.8m ()}, \facility{Lulin:1.8m ()},
\facility{Beijing:2.16m ()}, \facility{Maidanek:1.5m ()}

\bibliographystyle{apj}

\begin{thebibliography}{66}
\expandafter\ifx\csname natexlab\endcsname\relax\def\natexlab#1{#1}\fi

\bibitem[{{Althaus} {et~al.}(2009){Althaus}, {Panei}, {Romero}, {Rohrmann},
  {C{\'o}rsico}, {Garc{\'{\i}}a-Berro}, \& {Miller Bertolami}}]{Althaus09}
{Althaus}, L.~G., {Panei}, J.~A., {Romero}, A.~D., {Rohrmann}, R.~D.,
  {C{\'o}rsico}, A.~H., {Garc{\'{\i}}a-Berro}, E., \& {Miller Bertolami}, M.~M.
  2009, \aap, 502, 207

\bibitem[{{Althaus} {et~al.}(2005){Althaus}, {Serenelli}, {Panei},
  {C{\'o}rsico}, {Garc{\'{\i}}a-Berro}, \& {Sc{\'o}ccola}}]{Althaus05}
{Althaus}, L.~G., {Serenelli}, A.~M., {Panei}, J.~A., {C{\'o}rsico}, A.~H.,
  {Garc{\'{\i}}a-Berro}, E., \& {Sc{\'o}ccola}, C.~G. 2005, A\&A, 435, 631

\bibitem[{{Angulo} {et~al.}(1999){Angulo}, {Arnould}, {Rayet}, {Descouvemont},
  {Baye}, {Leclercq-Willain}, {Coc}, {Barhoumi}, {Aguer}, {Rolfs}, {Kunz},
  {Hammer}, {Mayer}, {Paradellis}, {Kossionides}, {Chronidou}, {Spyrou},
  {degl'Innocenti}, {Fiorentini}, {Ricci}, {Zavatarelli}, {Providencia},
  {Wolters}, {Soares}, {Grama}, {Rahighi}, {Shotter}, \& {Lamehi
  Rachti}}]{Angulo99}
{Angulo}, C., {et~al.} 1999, Nuclear Physics A, 656, 3

\bibitem[{{B\" ohm-Vitense}(1958)}]{Bohm-Vitense58}
{B\" ohm-Vitense}, E. 1958, Zeitschrift f\"ur Astrophysik, 46, 108

\bibitem[{{Beauchamp} {et~al.}(1999){Beauchamp}, {Wesemael}, {Bergeron},
  {Fontaine}, {Saffer}, {Liebert}, \& {Brassard}}]{Beauchamp99}
{Beauchamp}, A., {Wesemael}, F., {Bergeron}, P., {Fontaine}, G., {Saffer},
  R.~A., {Liebert}, J., \& {Brassard}, P. 1999, \apj, 516, 887

\bibitem[{{B{\'e}dard} {et~al.}(2017){B{\'e}dard}, {Bergeron}, \&
  {Fontaine}}]{Bedard17}
{B{\'e}dard}, A., {Bergeron}, P., \& {Fontaine}, G. 2017, \apj, 848, 11

\bibitem[{{Bell} {et~al.}(2015){Bell}, {Hermes}, {Bischoff-Kim}, {Moorhead},
  {Montgomery}, {{\O}stensen}, {Castanheira}, \& {Winget}}]{Bell15}
{Bell}, K.~J., {Hermes}, J.~J., {Bischoff-Kim}, A., {Moorhead}, S.,
  {Montgomery}, M.~H., {{\O}stensen}, R., {Castanheira}, B.~G., \& {Winget},
  D.~E. 2015, \apj, 809, 14

\bibitem[{{Bergeron} {et~al.}(2011){Bergeron}, {Wesemael}, {Dufour},
  {Beauchamp}, {Hunter}, {Saffer}, {Gianninas}, {Ruiz}, {Limoges}, {Dufour},
  {Fontaine}, \& {Liebert}}]{Bergeron11}
{Bergeron}, P., {et~al.} 2011, \apj, 737, 28

\bibitem[{{Bischoff-Kim}(2015)}]{Bischoff-Kim15}
{Bischoff-Kim}, A. 2015, in Astronomical Society of the Pacific Conference
  Series, Vol. 493, 19th European Workshop on White Dwarfs, ed. P.~{Dufour},
  P.~{Bergeron}, \& G.~{Fontaine}, 175

\bibitem[{{Bischoff-Kim} \& {Metcalfe}(2011)}]{Bischoff-Kim11a}
{Bischoff-Kim}, A., \& {Metcalfe}, T.~S. 2011, MNRAS, 414, 404

\bibitem[{{Bischoff-Kim} {et~al.}(2014){Bischoff-Kim}, {{\O}stensen}, {Hermes},
  \& {Provencal}}]{Bischoff-Kim14}
{Bischoff-Kim}, A., {{\O}stensen}, R.~H., {Hermes}, J.~J., \& {Provencal},
  J.~L. 2014, ApJ, 794, 39

\bibitem[{{Bischoff-Kim} \& {Provencal}(2017)}]{Bischoff-Kim17}
{Bischoff-Kim}, A., \& {Provencal}, J.~L. 2017, in 20th European White Dwarf
  Workshop, ed. P.~E. {Tremblay}, B.~{Gaensicke}, \& T.~{Marsh}, Vol. 509, 309

\bibitem[{{Bradley}(2004)}]{Bradley04}
{Bradley}, P.~A. 2004, in Astronomical Society of the Pacific Conference
  Series, Vol. 310, IAU Colloq. 193: Variable Stars in the Local Group, ed.
  D.~W. {Kurtz} \& K.~R. {Pollard}, 506

\bibitem[{{Bradley} \& {Winget}(1991)}]{Bradley91}
{Bradley}, P.~A., \& {Winget}, D.~E. 1991, \apjs, 75, 463

\bibitem[{{Bradley} \& {Winget}(1994)}]{Bradley94a}
---. 1994, \apj, 430, 850

\bibitem[{{Bradley} {et~al.}(1993){Bradley}, {Winget}, \& {Wood}}]{Bradley93}
{Bradley}, P.~A., {Winget}, D.~E., \& {Wood}, M.~A. 1993, \apj, 406, 661

\bibitem[{{Brassard} {et~al.}(1995){Brassard}, {Fontaine}, \&
  {Wesemael}}]{Brassard95}
{Brassard}, P., {Fontaine}, G., \& {Wesemael}, F. 1995, \apjs, 96, 545

\bibitem[{{Brickhill}(1992)}]{Brickhill92}
{Brickhill}, A.~J. 1992, \mnras, 259, 529

\bibitem[{{Castanheira} \& {Kepler}(2009)}]{Castanheira09}
{Castanheira}, B.~G., \& {Kepler}, S.~O. 2009, \mnras, 396, 1709

\bibitem[{{Castanheira} {et~al.}(2005){Castanheira}, {Nitta}, {Kepler},
  {Winget}, \& {Koester}}]{Castanheira05}
{Castanheira}, B.~G., {Nitta}, A., {Kepler}, S.~O., {Winget}, D.~E., \&
  {Koester}, D. 2005, \aap, 432, 175

\bibitem[{{Charpinet} {et~al.}(2008){Charpinet}, {Van Grootel}, {Reese},
  {Fontaine}, {Green}, {Brassard}, \& {Chayer}}]{Charpinet08}
{Charpinet}, S., {Van Grootel}, V., {Reese}, D., {Fontaine}, G., {Green},
  E.~M., {Brassard}, P., \& {Chayer}, P. 2008, \aap, 489, 377

\bibitem[{{Dalessio}(2010)}]{Dalessio2010}
{Dalessio}, J. 2010, American Astronomical Society Meeting Abstracts, 215,
  452.09

\bibitem[{{De Ger{\'o}nimo} {et~al.}(2017){De Ger{\'o}nimo}, {Althaus},
  {C{\'o}rsico}, {Romero}, \& {Kepler}}]{DeGeronimo17}
{De Ger{\'o}nimo}, F.~C., {Althaus}, L.~G., {C{\'o}rsico}, A.~H., {Romero},
  A.~D., \& {Kepler}, S.~O. 2017, \aap, 599, A21

\bibitem[{{deBoer} {et~al.}(2017){deBoer}, {G{\"o}rres}, {Wiescher}, {Azuma},
  {Best}, {Brune}, {Fields}, {Jones}, {Pignatari}, {Sayre}, {Smith}, {Timmes},
  \& {Uberseder}}]{deBoer17}
{deBoer}, R.~J., {et~al.} 2017, Reviews of Modern Physics, 89, 035007

\bibitem[{{Dehner} \& {Kawaler}(1995)}]{Dehner95}
{Dehner}, B.~T., \& {Kawaler}, S.~D. 1995, \apjl, 445, L141

\bibitem[{{Dreizler} \& {Heber}(1998)}]{Dreizler98}
{Dreizler}, S., \& {Heber}, U. 1998, A\&A, 334, 618

\bibitem[{{Fontaine} \& {Brassard}(2002)}]{Fontaine02}
{Fontaine}, G., \& {Brassard}, P. 2002, \apjl, 581, L33

\bibitem[{{Gaia Collaboration} {et~al.}(2018){Gaia Collaboration}, {Brown},
  {Vallenari}, {Prusti}, {de Bruijne}, {Babusiaux}, \& {Bailer-Jones}}]{Gaia18}
{Gaia Collaboration}, {Brown}, A.~G.~A., {Vallenari}, A., {Prusti}, T., {de
  Bruijne}, J.~H.~J., {Babusiaux}, C., \& {Bailer-Jones}, C.~A.~L. 2018, ArXiv
  e-prints

\bibitem[{{Giammichele} {et~al.}(2016){Giammichele}, {Fontaine}, {Brassard}, \&
  {Charpinet}}]{Giammichele16}
{Giammichele}, N., {Fontaine}, G., {Brassard}, P., \& {Charpinet}, S. 2016,
  \apjs, 223, 10

\bibitem[{{Giammichele} {et~al.}(2018){Giammichele}, {Charpinet}, {Fontaine},
  {Brassard}, {Green}, {Van Grootel}, {Bergeron}, {Zong}, \&
  {Dupret}}]{Giammichele18}
{Giammichele}, N., {et~al.} 2018, \nat, 554, 73

\bibitem[{{Hermes} {et~al.}(2017){Hermes}, {Kawaler}, {Bischoff-Kim},
  {Provencal}, {Dunlap}, \& {Clemens}}]{Hermes2017}
{Hermes}, J.~J., {Kawaler}, S.~D., {Bischoff-Kim}, A., {Provencal}, J.~L.,
  {Dunlap}, B.~H., \& {Clemens}, J.~C. 2017, \apj, 835, 277

\bibitem[{{Herwig} {et~al.}(1999){Herwig}, {Bl{\"o}cker}, {Langer}, \&
  {Driebe}}]{Herwig99}
{Herwig}, F., {Bl{\"o}cker}, T., {Langer}, N., \& {Driebe}, T. 1999, A\&A, 349,
  L5

\bibitem[{{Iben} {et~al.}(1983){Iben}, {Kaler}, {Truran}, \&
  {Renzini}}]{Iben83}
{Iben}, Jr., I., {Kaler}, J.~B., {Truran}, J.~W., \& {Renzini}, A. 1983, \apj,
  264, 605

\bibitem[{{Ising} \& {Koester}(2001)}]{Ising01}
{Ising}, J., \& {Koester}, D. 2001, \aap, 374, 116

\bibitem[{{Kepler} {et~al.}(2003){Kepler}, {Nather}, {Winget}, {Nitta},
  {Kleinman}, {Metcalfe}, {Sekiguchi}, {Xiaojun}, {Sullivan}, {Sullivan},
  {Janulis}, {Meistas}, {Kalytis}, {Krzesinski}, {Ogoza}, {Zola}, {O'Donoghue},
  {Romero-Colmenero}, {Martinez}, {Dreizler}, {Deetjen}, {Nagel}, {Schuh},
  {Vauclair}, {Ning}, {Chevreton}, {Solheim}, {Gonzalez Perez}, {Johannessen},
  {Kanaan}, {Costa}, {Murillo Costa}, {Wood}, {Silvestri}, {Ahrens}, {Jones},
  {Collins}, {Boyer}, {Shaw}, {Mukadam}, {Klumpe}, {Larrison}, {Kawaler},
  {Riddle}, {Ulla}, \& {Bradley}}]{Kepler03}
{Kepler}, S.~O., {et~al.} 2003, \aap, 401, 639

\bibitem[{{Kleinman} {et~al.}(1998){Kleinman}, {Nather}, {Winget}, {Clemens},
  {Bradley}, {Kanaan}, {Provencal}, {Claver}, {Watson}, {Yanagida}, {Nitta},
  {Dixson}, {Wood}, {Grauer}, {Hine}, {Fontaine}, {Liebert}, {Sullivan},
  {Wickramasinghe}, {Achilleos}, {Marar}, {Seetha}, {Ashoka}, {Meistas},
  {Leibowitz}, {Moskalik}, {Krzesinzski}, {Solheim}, {Bruvold}, {O'Donoghue},
  D., {Warner}, {Martinez}, {Vauclair}, {Dolez}, {Chevreton}, {Barstow},
  {Kepler}, {Giovannini}, {Augusteijn}, {Hansen}, \& {Kawaler}}]{Kleinman98}
{Kleinman}, S.~J., {et~al.} 1998, \apj, 495, 424

\bibitem[{{Koester}(2013)}]{Koester2013}
{Koester}, D. 2013, private communication

\bibitem[{{Koester} {et~al.}(2014){Koester}, {Provencal}, \&
  {G{\"a}nsicke}}]{Koester2014}
{Koester}, D., {Provencal}, J., \& {G{\"a}nsicke}, B.~T. 2014, \aap, 568, A118

\bibitem[{{Kotak} {et~al.}(2002){Kotak}, {van Kerkwijk}, \&
  {Clemens}}]{Kotak02}
{Kotak}, R., {van Kerkwijk}, M.~H., \& {Clemens}, J.~C. 2002, \aap, 388, 219

\bibitem[{{Lamb} \& {van Horn}(1975)}]{Lamb75}
{Lamb}, D.~Q., \& {van Horn}, H.~M. 1975, \apj, 200, 306

\bibitem[{{Lawlor} \& {MacDonald}(2006)}]{Lawlor06}
{Lawlor}, T.~M., \& {MacDonald}, J. 2006, \mnras, 371, 263

\bibitem[{{Lenz} \& {Breger}(2005)}]{Lenz05}
{Lenz}, P., \& {Breger}, M. 2005, Communications in Asteroseismology, 146, 53

\bibitem[{{Metcalfe}(2003)}]{Metcalfe03}
{Metcalfe}, T.~S. 2003, \apjl, 587, L43

\bibitem[{{Metcalfe} {et~al.}(2003{\natexlab{a}}){Metcalfe}, {Montgomery}, \&
  {Kanaan}}]{Metcalfe03c}
{Metcalfe}, T.~S., {Montgomery}, M.~H., \& {Kanaan}, A. 2003{\natexlab{a}},
  American Astronomical Society Meeting, 203

\bibitem[{{Metcalfe} {et~al.}(2005){Metcalfe}, {Montgomery}, \&
  {Kanaan}}]{Metcalfe05a}
{Metcalfe}, T.~S., {Montgomery}, M.~H., \& {Kanaan}, A. 2005, in Astronomical
  Society of the Pacific Conference Series, Vol. 334, 14th European Workshop on
  White Dwarfs, ed. D.~{Koester} \& S.~{Moehler}, 465

\bibitem[{{Metcalfe} {et~al.}(2003{\natexlab{b}}){Metcalfe}, {Montgomery}, \&
  {Winget}}]{Metcalfe03b}
{Metcalfe}, T.~S., {Montgomery}, M.~H., \& {Winget}, D.~E. 2003{\natexlab{b}},
  White Dwarfs: Galactic and Cosmologic Probes, 25th meeting of the IAU, Joint
  Discussion 5, 16-17 July 2003, Sydney, Australia, 5

\bibitem[{{Metcalfe} {et~al.}(2000){Metcalfe}, {Nather}, \&
  {Winget}}]{Metcalfe00}
{Metcalfe}, T.~S., {Nather}, R.~E., \& {Winget}, D.~E. 2000, \apj, 545, 974

\bibitem[{{Metcalfe} {et~al.}(2002){Metcalfe}, {Salaris}, \&
  {Winget}}]{Metcalfe02}
{Metcalfe}, T.~S., {Salaris}, M., \& {Winget}, D.~E. 2002, \apj, 573, 803

\bibitem[{{Metcalfe} {et~al.}(2001){Metcalfe}, {Winget}, \&
  {Charbonneau}}]{Metcalfe01}
{Metcalfe}, T.~S., {Winget}, D.~E., \& {Charbonneau}, P. 2001, \apj, 557, 1021

\bibitem[{{Montgomery} {et~al.}(2016){Montgomery}, {Hermes}, {Dunlap},
  {Winget}, {Bell}, {Provencal}, {Clemens}, \& {Fanale}}]{Mont2016}
{Montgomery}, M.~H., {Hermes}, J.~J., {Dunlap}, B.~H., {Winget}, D.~E., {Bell},
  K.~J., {Provencal}, J.~L., {Clemens}, J.~C., \& {Fanale}, S. 2016, private
  communication

\bibitem[{{Montgomery} {et~al.}(2003){Montgomery}, {Metcalfe}, \&
  {Winget}}]{Montgomery03}
{Montgomery}, M.~H., {Metcalfe}, T.~S., \& {Winget}, D.~E. 2003, \mnras, 344,
  657

\bibitem[{{Montgomery} {et~al.}(2010){Montgomery}, {Provencal}, {Kanaan},
  {Thompson}, {Dalessio}, {Shipman}, \& {Winget}}]{Montgomery10}
{Montgomery}, M.~H., {Provencal}, J.~L., {Kanaan}, A., {Thompson}, S.,
  {Dalessio}, J., {Shipman}, H., \& {Winget}, D.~E. 2010, \apj, 716, 84

\bibitem[{{Nather} {et~al.}(1990){Nather}, {Winget}, {Clemens}, {Hansen}, \&
  {Hine}}]{wet90}
{Nather}, R.~E., {Winget}, D.~E., {Clemens}, J.~C., {Hansen}, C.~J., \& {Hine},
  B.~P. 1990, \apj, 361, 309

\bibitem[{{Nitta} {et~al.}(2012){Nitta}, {Koester}, {Chu}, {Thompson},
  {Kepler}, {Kleinman}, {Winget}, {Provencal}, \& {Castanheira}}]{Nitta12}
{Nitta}, A., {et~al.} 2012, in Astronomical Society of the Pacific Conference
  Series, Vol. 462, Progress in Solar/Stellar Physics with Helio- and
  Asteroseismology, ed. H.~{Shibahashi}, M.~{Takata}, \& A.~E. {Lynas-Gray},
  171

\bibitem[{{Provencal} {et~al.}(2009){Provencal}, {Montgomery}, {Kanaan},
  {Shipman}, {Childers}, {Baran}, {Kepler}, {Reed}, {Zhou}, {Eggen}, {Watson},
  {Winget}, {Thompson}, {Riaz}, {Nitta}, {Kleinman}, {Crowe}, {Slivkoff},
  {Sherard}, {Purves}, {Binder}, {Knight}, {Kim}, {Chen}, {Yang}, {Lin}, {Lin},
  {Chen}, {Jiang}, {Sergeev}, {Mkrtichian}, {Andreev}, {Janulis}, {Siwak},
  {Zola}, {Koziel}, {Stachowski}, {Paparo}, {Bognar}, {Handler}, {Lorenz},
  {Steininger}, {Beck}, {Nagel}, {Kusterer}, {Hoffman}, {Reiff}, {Kowalski},
  {Vauclair}, {Charpinet}, {Chevreton}, {Solheim}, {Pakstiene}, {Fraga}, \&
  {Dalessio}}]{Provencal09}
{Provencal}, J.~L., {et~al.} 2009, \apj, 693, 564

\bibitem[{{Provencal} {et~al.}(2012){Provencal}, {Montgomery}, {Kanaan},
  {Thompson}, {Dalessio}, {Shipman}, {Childers}, {Clemens}, {Rosen},
  {Henrique}, {Bischoff-Kim}, {Strickland}, {Chandler}, {Walter}, {Watson},
  {Castanheira}, {Wang}, {Handler}, {Wood}, {Vennes}, {Nemeth}, {Kepler},
  {Reed}, {Nitta}, {Kleinman}, {Brown}, {Kim}, {Sullivan}, {Chen}, {Yang},
  {Shih}, {Jiang}, {Sergeev}, {Maksim}, {Janulis}, {Baliyan}, {Vats}, {Zola},
  {Baran}, {Winiarski}, {Ogloza}, {Paparo}, {Bognar}, {Papics}, {Kilkenny},
  {Sefako}, {Buckley}, {Loaring}, {Kniazev}, {Silvotti}, {Galleti}, {Nagel},
  {Vauclair}, {Dolez}, {Fremy}, {Perez}, {Almenara}, \& {Fraga}}]{Provencal12}
---. 2012, ApJ, 751, 91

\bibitem[{{Salaris} {et~al.}(1997){Salaris}, {Dominguez}, {Garcia-Berro},
  {Hernanz}, {Isern}, \& {Mochkovitch}}]{Salaris97}
{Salaris}, M., {Dominguez}, I., {Garcia-Berro}, E., {Hernanz}, M., {Isern}, J.,
  \& {Mochkovitch}, R. 1997, \apj, 486, 413

\bibitem[{{Scargle}(1982)}]{Scargle82}
{Scargle}, J.~D. 1982, \apj, 263, 835

\bibitem[{{Sullivan} {et~al.}(2008){Sullivan}, {Metcalfe}, {O'Donoghue},
  {Winget}, {Kilkenny}, {van Wyk}, {Kanaan}, {Kepler}, {Nitta}, {Kawaler},
  {Montgomery}, {Nather}, {O'Brien}, {Bischoff-Kim}, {Wood}, {Jiang},
  {Leibowitz}, {Ibbetson}, {Zola}, {Krzesinski}, {Pajdosz}, {Vauclair},
  {Dolez}, \& {Chevreton}}]{Sullivan08}
{Sullivan}, D.~J., {et~al.} 2008, \mnras, 387, 137

\bibitem[{{Thompson} \& {Mullally}(2009)}]{wqed}
{Thompson}, S.~E., \& {Mullally}, F. 2009, Journal of Physics Conference
  Series, 172, 012081

\bibitem[{{Tremblay} {et~al.}(2013){Tremblay}, {Ludwig}, {Steffen}, \&
  {Freytag}}]{Tremblay2013}
{Tremblay}, P.-E., {Ludwig}, H.-G., {Steffen}, M., \& {Freytag}, B. 2013, \aap,
  559, A104

\bibitem[{{Unno} {et~al.}(1989){Unno}, {Osaki}, {Ando}, {Saio}, \&
  {Shibahashi}}]{Unno89}
{Unno}, W., {Osaki}, Y., {Ando}, H., {Saio}, H., \& {Shibahashi}, H. 1989,
  Nonradial Oscillations of Stars (Tokyo: University of Tokyo Press)

\bibitem[{{Winget} {et~al.}(1982){Winget}, {van Horn}, {Tassoul}, {Fontaine},
  {Hansen}, \& {Carroll}}]{Winget82}
{Winget}, D.~E., {van Horn}, H.~M., {Tassoul}, M., {Fontaine}, G., {Hansen},
  C.~J., \& {Carroll}, B.~W. 1982, \apjl, 252, L65

\bibitem[{{Winget} {et~al.}(1994){Winget}, {Nather}, {Clemens}, {Provencal},
  {Kleinman}, {Bradley}, {Claver}, {Dixson}, {Montgomery}, {Hansen}, {Hine},
  {Birch}, {Candy}, {Marar}, {Seetha}, {Ashoka}, {Leibowitz}, {O'Donoghue},
  {Warner}, {Buckley}, {Tripe}, {Vauclair}, {Dolez}, {Chevreton}, {Serre},
  {Garrido}, {Kepler}, {Kanaan}, {Augusteijn}, {Wood}, {Bergeron}, \&
  {Grauer}}]{Winget94}
{Winget}, D.~E., {et~al.} 1994, \apj, 430, 839

\bibitem[{{Wood}(1990)}]{Wood90}
{Wood}, M.~A. 1990, PhD thesis, The University of Texas at Austin

\bibitem[{{Wu}(2001)}]{Wu01}
{Wu}, Y. 2001, \mnras, 323, 248

\end{thebibliography}

\clearpage

\begin{deluxetable}{llrrr}
\tablecolumns{5}
\tabletypesize{\small}
\tablewidth{0pc}
\tablecaption{\label{journal}}
\tablehead{
\colhead{Run Name} & \colhead{Telescope} & \colhead{Detector} & \colhead{Date} & 
\colhead{Length}\\
\colhead{} & \colhead{} & \colhead{} & \colhead{} & \colhead{(hrs)}
}
\startdata
mcao070524-01 & MCAO 0.6 & CCD & 2007-05-24 & 4.0\\
mcao070531-02 & MCAO 0.6 & CCD & 2007-05-31 & 4.1\\
mcao080613-01 & MCAO 0.6 & CCD & 2008-06-13 & 2.0\\
mcao080617-02 & MCAO 0.6 & CCD & 2008-07-25 & 1.2\\
mcao080726-01 & MCAO 0.6 & CCD & 2008-07-26 & 1.7\\
mcao080730-01 & MCAO 0.6 & CCD & 2008-07-30 & 3.5\\
mcao080802-01 & MCAO 0.6 & CCD & 2008-08-02 & 3.7\\
pjmo080706-03 & PJMO 0.6 & CCD & 2008-07-06 & 5.1\\
suho080810-19 & Suhora 0.6 & CCD & 2008-08-10 & 4.7\\
suho080811-19 & Suhora 0.6 & CCD & 2008-08-11 & 2.5\\
boao090528-17 & BOAO 1.8 & CCD & 2009-05-28 & 2.0\\
kore090529-16 & BOAO 1.8 & CCD & 2009-05-29 & 2.4\\
mcao090513-01 & MCAO 0.6 & CCD & 2009-05-13 & 2.4\\
mcao090519-03 & MCAO 0.6 & CCD & 2009-05-19 & 3.2\\
mcao090520-01 & MCAO 0.6 & CCD & 2009-05-20 & 4.8\\
mcao090521-01 & MCAO 0.6 & CCD & 2009-05-21 & 4.9\\
mcao090522-01 & MCAO 0.6 & CCD & 2009-05-22 & 3.1\\
mcao090531-02 & MCAO 0.6 & CCD & 2009-05-31 & 2.1\\
mcao090601-01 & MCAO 0.6 & CCD & 2009-06-01 & 3.3\\
mole090526-20 & Moletai 1.65  & CCD & 2009-05-26 & 3.7\\
suho090520-19 & Suhora 0.6  & CCD & 2009-05-20 & 2.2\\
suho090521-19 & Suhora 0.6 & CCD & 2009-05-21 & 5.1\\
vien090525-19 & Vienna 0.6 & CCD & 2009-05-25 & 6.2\\
krak100523-20 & Krakow 0.4 & CCD & 2010-05-23 & 3.4\\
krak100526-20 & Krakow 0.4 & CCD & 2010-05-26 & 2.3\\
krak100616-20 & Krakow 0.4 & CCD & 2010-06-16 & 4.3\\
krak100617-20 & Krakow 0.4 & CCD & 2010-06-17 & 4.5\\
mcao100516-01 & MCAO 0.6 & CCD & 2010-05-16 & 2.0\\
mcao100521-01 & MCAO 0.6 & CCD & 2010-05-21 & 0.9\\
mcao100521-05 & MCAO 0.6 & CCD & 2010-05-21 & 0.8\\
mcao100526-01 & MCAO 0.6 & CCD & 2010-05-26 & 3.3\\
mcdo100517-06 & McDonald 2.1  & CCD & 2010-05-17 & 4.8\\
mole100516-22 & Moletai 1.65  & CCD & 2010-05-26 & 1.8\\
mole100517-22 & Moletai 1.65  & CCD & 2010-05-17 & 2.0\\
mole100520-21 & Moletai 1.65  & CCD & 2010-05-20 & 3.3\\
mole100521-21 & Moletai 1.65  & CCD & 2010-05-21 & 3.5\\
pjmo100519-04 & PJMO 0.6 & CCD & 2010-05-19 & 4.0\\
pjmo100520-02 & PJMO 0.6 & CCD & 2010-05-20 & 3.8\\
pjmo100521-05 & PJMO 0.6 & CCD & 2010-05-21 & 3.3\\
pjmo100522-02 & PJMO 0.6 & CCD & 2010-05-22 & 7.0\\
pjmo100523-04 & PJMO 0.6 & CCD & 2010-05-23 & 6.0\\
pjmo100524-02 & PJMO 0.6 & CCD & 2010-05-24 & 7.5\\
pjmo100527-02 & PJMO 0.6 & CCD & 2010-05-27 & 8.3\\
pjmo100528-02 & PJMO 0.6 & CCD & 2010-05-28 & 8.3\\
pjmo100619-03 & PJMO 0.6 & CCD & 2010-06-19 & 4.0\\
pjmo100622-02 & PJMO 0.6 & CCD & 2010-06-22 & 0.7\\
suho100617-21 & Suhora 0.6 & CCD & 2010-06-17 & 4.3\\
terb100516-17 & Terskol 2.0  & CCD & 2010-05-16 & 2.4\\
terb100520-17 & Terskol 2.0  & CCD & 2010-05-20 & 8.9\\
terb100521-20 & Terskol 2.0  & CCD & 2010-05-21 & 4.0\\
terb100524-19 & Terskol 2.0  & CCD & 2010-05-24 & 5.6\\
terb100525-17 & Terskol 2.0  & CCD & 2010-05-25 & 7.3\\
tueb100522-20 & Tuebingen 0.8 & SBig & 2010-05-22 & 5.9\\
tueb100523-20 & Tuebingen 0.8 & SBig & 2010-05-23 & 5.9\\
tueb100524-20 & Tuebingen 0.8 & SBig & 2010-05-24 & 6.0\\
tueb100604-20 & Tuebingen 0.8 & SBig & 2010-06-04 & 5.8\\
tueb100605-20 & Tuebingen 0.8 & SBig & 2010-06-05 & 5.3\\
turk100512-18 & Canakkale 1.2 & CCD & 2010-05-12 & 6.6\\
turk100513-22 & Canakkale 1.2 & CCD & 2010-05-13 & 2.8\\
turk100516-19 & Canakkale 1.2 & CCD & 2010-05-16 & 6.4\\
turk100520-19 & Canakkale 1.2 & CCD & 2010-05-20 & 1.1\\
turk100522-21 & Canakkale 1.2 & CCD & 2010-05-22 & 2.6\\
turk100523-19 & Canakkale 1.2 & CCD & 2010-05-23 & 6.0\\
turk100611-19 & Canakkale 1.2 & CCD & 2010-06-11 & 5.2\\
turk100613-19 & Canakkale 1.2 & CCD & 2010-06-13 & 5.3\\
turk100615-20 & Canakkale 1.2 & CCD & 2010-06-15 & 4.8\\
turk100618-19 & Canakkale 1.2 & CCD & 2010-06-18 & 4.5\\
turk100620-20 & Canakkale 1.2 & CCD & 2010-06-20 & 4.8\\
hvar110521-22 & Hvar 1.0 & CCD & 2011-05-21 & 2.6\\
hvar110526-19 & Hvar 1.0 & CCD & 2011-05-26 & 6.6\\
hvar110527-19 & Hvar 1.0 & CCD & 2011-05-27 & 3.5\\
hvar110529-19 & Hvar 1.0 & CCD & 2011-05-29 & 6.7\\
hvar110531-19 & Hvar 1.0 & CCD & 2011-05-31 & 4.3\\
hvar110602-19 & Hvar 1.0 & CCD & 2011-06-02 & 3.8\\
hvar110604-19 & Hvar 1.0 & CCD & 2011-06-04 & 6.6\\
hvar110606-20 & Hvar 1.0 & CCD & 2011-06-06 & 4.4\\
krak110506-20 & Krakow 0.4 & CCD & 2011-05-06 & 5.4\\
krak110509-19 & Krakow 0.4 & CCD & 2011-05-09 & 6.7\\
krak110510-19 & Krakow 0.4 & CCD & 2011-05-10 & 6.2\\
krak110511-19 & Krakow 0.4 & CCD & 2011-05-11 & 6.7\\
krak110512-19 & Krakow 0.4 & CCD & 2011-05-12 & 3.1\\
krak110516-19 & Krakow 0.4 & CCD & 2011-05-16 & 6.1\\
krak110517-19 & Krakow 0.4 & CCD & 2011-05-17 & 5.6\\
krak110518-19 & Krakow 0.4 & CCD & 2011-05-18 & 5.7\\
krak110519-20 & Krakow 0.4 & CCD & 2011-05-19 & 5.4\\
krak110520-20 & Krakow 0.4 & CCD & 2011-05-20 & 5.2\\
krak110522-21 & Krakow 0.4 & CCD & 2011-05-22 & 4.3\\
krak110523-23 & Krakow 0.4 & CCD & 2011-05-23 & 1.4\\
krak110524-19 & Krakow 0.4 & CCD & 2011-05-24 & 4.0\\
krak110525-20 & Krakow 0.4 & CCD & 2011-05-25 & 5.4\\
krak110526-20 & Krakow 0.4 & CCD & 2011-05-26 & 5.3\\
krak110529-20 & Krakow 0.4 & CCD & 2011-05-29 & 5.0\\
krak110530-20 & Krakow 0.4 & CCD & 2011-05-30 & 4.8\\
krak110531-20 & Krakow 0.4 & CCD & 2011-05-31 & 0.8\\
krak110604-20 & Krakow 0.4 & CCD & 2011-06-04 & 5.3\\
mole110426-22 & Moletai 1.65  & CCD & 2011-04-26 & 1.7\\
mole110427-21 & Moletai 1.65  & CCD & 2011-04-27 & 3.7\\
mole110430-20 & Moletai 1.65  & CCD & 2011-04-30 & 0.8\\
mole110502-20 & Moletai 1.65  & CCD & 2011-05-02 & 0.4\\
mole110502-21 & Moletai 1.65  & CCD & 2011-05-02 & 2.6\\
mole110505-20 & Moletai 1.65  & CCD & 2011-05-05 & 1.1\\
mole110510-21 & Moletai 1.65  & CCD & 2011-05-10 & 3.5\\
mtlm110426-05 & Mt. Lemmon 1.0 & CCD & 2011-04-26 & 6.3\\
mtlm110427-06 & Mt. Lemmon 1.0 & CCD & 2011-04-27 & 5.5\\
mtlm110428-04 & Mt. Lemmon 1.0 & CCD & 2011-04-28 & 7.1\\
mtlm110429-05 & Mt. Lemmon 1.0 & CCD & 2011-04-29 & 6.8\\
mtlm110430-05 & Mt. Lemmon 1.0 & CCD & 2011-04-30 & 6.4\\
mtlm110501-05 & Mt. Lemmon 1.0 & CCD & 2011-05-01 & 6.6\\
mtlm110502-05 & Mt. Lemmon 1.0 & CCD & 2011-05-02 & 6.5\\
naoc110426-13 & NAOC 0.5 & CCD & 2011-04-26 & 6.9\\
naoc110427-11 & NAOC 0.5 & CCD & 2011-04-27 & 3.3\\
naos110426-13 & NAOC 0.85 & CCD & 2011-04-26 & 6.9\\
naos110427-11 & NAOC 0.85 & CCD & 2011-04-27 & 3.3\\
naos110428-12 & NAOC 0.85 & CCD & 2011-04-28 & 2.6\\
naos110501-13 & NAOC 0.85 & CCD & 2011-05-01 & 7.1\\
naos110502-12 & NAOC 0.85 & CCD & 2011-05-02 & 8.3\\
naos110505-13 & NAOC 0.85 & CCD & 2011-05-05 & 7.0\\
pjmo110426-07 & PJMO 0.6 & CCD & 2011-04-26 & 3.0\\
pjmo110428-04 & PJMO 0.6 & CCD & 2011-04-28 & 2.0\\
pjmo110429-03 & PJMO 0.6 & CCD & 2011-04-29 & 2.5\\
pjmo110430-04 & PJMO 0.6 & CCD & 2011-04-30 & 3.0\\
pjmo110503-04 & PJMO 0.6 & CCD & 2011-05-03 & 2.6\\
pjmo110518-03 & PJMO 0.6 & CCD & 2011-05-18 & 6.7\\
pjmo110519-03 & PJMO 0.6 & CCD & 2011-05-19 & 2.7\\
pjmo110521-03 & PJMO 0.6 & CCD & 2011-05-21 & 2.7\\
pjmo110524-02 & PJMO 0.6 & CCD & 2011-05-24 & 5.2\\
pjmo110525-07 & PJMO 0.6 & CCD & 2011-05-25 & 3.2\\
pjmo110526-03 & PJMO 0.6 & CCD & 2011-05-26 & 7.0\\
pjmo110527-02 & PJMO 0.6 & CCD & 2011-05-27 & 0.8\\
pjmo110527-03 & PJMO 0.6 & CCD & 2011-05-27 & 6.7\\
suho110521-20 & Suhora 0.6 & CCD & 2011-05-21 & 5.2\\
suho110522-19 & Suhora 0.6 & CCD & 2011-05-22 & 5.9\\
suho110523-20 & Suhora 0.6 & CCD & 2011-05-23 & 5.6\\
suho110527-20 & Suhora 0.6 & CCD & 2011-05-27 & 2.0\\
suho110529-20 & Suhora 0.6 & CCD & 2011-05-29 & 4.6\\
suho110530-20 & Suhora 0.6 & CCD & 2011-05-30 & 4.9\\
terb110527-20 & Terskol 2.0  & CCD & 2011-05-27 & 2.4\\
terb110529-17 & Terskol 2.0  & CCD & 2011-05-27 & 4.4\\
terb110530-18 & Terskol 2.0  & CCD & 2011-05-30 & 2.6\\
terb110531-17 & Terskol 2.0  & CCD & 2011-05-31 & 1.8\\
terb110601-19 & Terskol 2.0  & CCD & 2011-06-01 & 3.3\\
tueb110504-19 & Tuebingen 0.8 & SBig & 2011-05-04 & 6.8\\
tueb110505-19 & Tuebingen 0.8 & SBig & 2011-05-05 & 6.8\\
tueb110506-19 & Tuebingen 0.8 & SBig & 2011-05-06 & 6.7\\
tueb110508-19 & Tuebingen 0.8 & SBig & 2011-05-08 & 6.6\\
tueb110509-19 & Tuebingen 0.8 & SBig & 2011-05-09 & 6.7\\
tueb110513-19 & Tuebingen 0.8 & SBig & 2011-05-13 & 2.2\\
tueb110518-22 & Tuebingen 0.8 & SBig & 2011-05-18 & 4.1\\
tueb110523-20 & Tuebingen 0.8 & SBig & 2011-05-23 & 5.9\\
tueb110524-20 & Tuebingen 0.8 & SBig & 2011-05-24 & 5.9\\
tueb110525-20 & Tuebingen 0.8 & SBig & 2011-05-24 & 6.0\\
tueb110529-20 & Tuebingen 0.8 & SBig & 2011-05-29 & 5.9\\
tueb110530-20 & Tuebingen 0.8 & SBig & 2011-05-30 & 3.4\\
tubi110602-00 & Tubitak 1.0  & CCD & 2011-06-02 & 1.1\\
tubi110602-23 & Tubitak 1.0  & CCD & 2011-06-02 & 2.0\\
tubi110603-20 & Tubitak 1.0  & CCD & 2011-06-03 & 5.2\\
boao120418-18 & BOAO 1.8 & CCD & 2012-04-18 & 1.7\\
krak120418-00 & Krakow 0.4 & CCD & 2012-04-18 & 1.8\\
krak120421-00 & Krakow 0.4 & CCD & 2012-04-21 & 4.6\\
krak120423-00 & Krakow 0.4 & CCD & 2012-04-23 & 1.7\\
krak120428-00 & Krakow 0.4 & CCD & 2012-04-28 & 6.6\\
krak120429-00 & Krakow 0.4 & CCD & 2012-04-29 & 6.1\\
krak120430-00 & Krakow 0.4 & CCD & 2012-04-30 & 6.7\\
mtlm120419-10 & Mt. Lemmon 1.0 & CCD & 2012-04-19 & 1.7\\
mtlm120420-10 & Mt. Lemmon 1.0 & CCD & 2012-04-20 & 1.4\\
mtlm120421-05 & Mt. Lemmon 1.0 & CCD & 2012-04-21 & 6.2\\
mtlm120422-07 & Mt. Lemmon 1.0 & CCD & 2012-04-22 & 4.8\\
na50120524-12 & NAOC 0.5 & CCD & 2012-05-24 & 4.5\\
na50120525-12 & NAOC 0.5 & CCD & 2012-05-25 & 4.4\\
na50120526-14 & NAOC 0.5 & CCD & 2012-05-26 & 3.3\\
na50120527-12 & NAOC 0.5 & CCD & 2012-05-27 & 4.9\\
na50120530-13 & NAOC 0.5 & CCD & 2012-05-30 & 6.0\\
pjmo120425-03 & PJMO 0.6 & CCD & 2012-04-25 & 7.2\\
pjmo120426-03 & PJMO 0.6 & CCD & 2012-04-26 & 4.3\\
prom120430-04 & PROMPT 0.4 & CCD & 2012-04-30 & 2.2\\
prom120430-07 & PROMPT 0.4 & CCD & 2012-04-30 & 1.1\\
prom120501-04 & PROMPT 0.4 & CCD & 2012-05-01 & 4.5\\
prom120502-04 & PROMPT 0.4 & CCD & 2012-05-02 & 4.4\\
prom120503-04 & PROMPT 0.4 & CCD & 2012-05-03 & 0.9\\
prom120504-04 & PROMPT 0.4 & CCD & 2012-05-04 & 4.5\\
prom120509-04 & PROMPT 0.4 & CCD & 2012-05-09 & 4.3\\
prom120510-04 & PROMPT 0.4 & CCD & 2012-05-10 & 4.1\\
prom120511-03 & PROMPT 0.4 & CCD & 2012-05-11 & 4.1\\
prom120512-03 & PROMPT 0.4 & CCD & 2012-05-12 & 4.0\\
prom120513-03 & PROMPT 0.4 & CCD & 2012-05-13 & 2.1\\
prom120514-03 & PROMPT 0.4 & CCD & 2012-05-14 & 4.0\\
prom120515-03 & PROMPT 0.4 & CCD & 2012-05-15 & 4.0\\
suho120501-20 & Suhora 0.6 & CCD & 2012-05-01 & 6.0\\
suho120502-21 & Suhora 0.6 & CCD & 2012-05-02 & 5.1\\
suho120503-23 & Suhora 0.6 & CCD & 2012-05-03 & 2.4\\
suho120505-19 & Suhora 0.6 & CCD & 2012-05-05 & 2.1\\
tubi120420-01 & Tubitak 1.0  & CCD & 2012-04-20 & 1.0\\
tubi120423-19 & Tubitak 1.0  & CCD & 2012-04-23 & 6.6\\
tubi120512-20 & Tubitak 1.0  & CCD & 2012-04-23 & 5.2\\
caam130412-21 & Cannakkale 1.2 & CCD & 2013-04-12 & 4.0\\
caam130417-21 & Cannakkale 1.2 & CCD & 2013-04-17 & 3.2\\
caam130418-21 & Cannakkale 1.2 & CCD & 2013-04-18 & 4.5\\
krak130418-00 & Krakow 0.4 & CCD & 2013-04-18 & 6.7\\
krak130421-00 & Krakow 0.4 & CCD & 2013-04-18 & 7.4\\
krak130425-00 & Krakow 0.4 & CCD & 2013-04-25 & 6.7\\
krak130426-00 & Krakow 0.4 & CCD & 2013-04-26 & 2.1\\
naos130503-13 & NAOC 0.85 & CCD & 2013-05-03 & 6.0\\
pjmo130429-06 & PJMO 0.6 & CCD & 2013-04-29 & 2.7\\
pjmo130430-07 & PJMO 0.6 & CCD & 2013-04-30 & 2.1\\
pjmo130503-07 & PJMO 0.6 & CCD & 2013-05-03 & 3.6\\
pjmo130505-04 & PJMO 0.6 & CCD & 2013-05-05 & 6.8\\
suho130422-21 & Suhora 0.6 & CCD & 2013-04-22 & 4.6\\
suho130425-20 & Suhora 0.6 & CCD & 2013-04-25 & 5.9\\
suho130426-20 & Suhora 0.6 & CCD & 2013-04-26 & 5.4\\
suho130501-19 & Suhora 0.6 & CCD & 2013-05-01 & 6.1\\
suho130504-21 & Suhora 0.6 & CCD & 2013-05-04 & 4.4\\
suho130505-19 & Suhora 0.6 & CCD & 2013-05-05 & 5.5\\
caam140610-19 & Cannakkale 1.2 & CCD & 2014-046-10 & 5.5\\
krak140530-21 & Krakow 0.4 & CCD & 2014-05-30 & 2.4\\
krak140604-20 & Krakow 0.4 & CCD & 2014-06-04 & 5.5\\
krak140606-20 & Krakow 0.4 & CCD & 2014-06-06 & 4.7\\
krak140607-20 & Krakow 0.4 & CCD & 2014-06-07 & 5.1\\
krak140608-20 & Krakow 0.4 & CCD & 2014-06-08 & 5.2\\
krak140609-19 & Krakow 0.4 & CCD & 2014-06-09 & 4.7\\
krak140610-20 & Krakow 0.4 & CCD & 2014-06-10 & 2.4\\
mcao140602-05 & MCAO 0.6 & CCD & 2014-06-02 & 3.0\\
mcao140603-01 & MCAO 0.6 & CCD & 2014-06-03 & 4.1\\
mcao140607-01 & MCAO 0.6 & CCD & 2014-06-07 & 4.0\\
mcao140608-01 & MCAO 0.6 & CCD & 2014-06-08 & 3.9\\
mole140526-20 & Moletai 1.65  & CCD & 2014-05-26 & 2.5\\
mole140527-20 & Moletai 1.65  & CCD & 2014-05-27 & 3.7\\
mole140605-20 & Moletai 1.65  & CCD & 2014-06-05 & 3.3\\
mole140607-20 & Moletai 1.65  & CCD & 2014-06-07 & 2.9\\
naos140605-15 & NAOC 0.85 & CCD & 2014-06-05 & 5.7\\
naos140606-17 & NAOC 0.85 & CCD & 2014-06-06 & 3.9\\
pjmo140519-04 & PJMO 0.6 & CCD & 2014-05-19 & 4.0\\
pjmo140529-03 & PJMO 0.6 & CCD & 2014-05-29 & 5.4\\
pjmo140530-03 & PJMO 0.6 & CCD & 2014-05-30 & 4.3\\
pjmo140531-02 & PJMO 0.6 & CCD & 2014-05-31 & 5.0\\
pjmo140601-03 & PJMO 0.6 & CCD & 2014-06-01 & 3.3\\
pjmo140602-02 & PJMO 0.6 & CCD & 2014-06-02 & 4.5\\
pjmo140603-02 & PJMO 0.6 & CCD & 2014-06-03 & 7.1\\
pjmo140604-04 & PJMO 0.6 & CCD & 2014-06-04 & 5.8\\
pjmo140611-03 & PJMO 0.6 & CCD & 2014-06-11 & 5.9\\
suho140604-19 & Suhora 0.6 & CCD & 2014-06-04 & 4.4\\
suho140606-20 & Suhora 0.6 & CCD & 2014-06-06 & 4.7\\
ters140604-20 & Terskol 0.6  & CCD & 2014-06-04 & 2.0\\
ters140619-20 & Terskol 0.6  & CCD & 2014-06-19 & 6.0\\
tsao140524-17 & Tien Shan 1.0 & CCD & 2014-05-24 & 4.7\\
tsao140525-16 & Tien Shan 1.0 & CCD & 2014-05-25 & 6.0\\
tsao140610-15 & Tien Shan 1.0 & CCD & 2014-06-10 & 1.0\\
tsao140611-16 & Tien Shan 1.0 & CCD & 2014-06-11 & 6.0\\
tsao140613-17 & Tien Shan 1.0 & CCD & 2014-06-13 & 4.6\\
tsao140616-19 & Tien Shan 1.0 & CCD & 2014-06-16 & 1.4\\
tsao140619-17 & Tien Shan 1.0 & CCD & 2014-06-19 & 4.8\\
tsao140620-18 & Tien Shan 1.0 & CCD & 2014-06-20 & 3.1\\
tsao140627-16 & Tien Shan 1.0 & CCD & 2014-06-27 & 5.9\\
tsao140628-16 & Tien Shan 1.0 & CCD & 2014-06-28 & 5.9\\
krak150420-20 & Krakow 0.4 & CCD & 2015-05-20 & 2.1\\
krak150421-19 & Krakow 0.4 & CCD & 2015-05-21 & 7.5\\
pjmo150420-03 & PJMO 0.6 & CCD & 2015-04-20 & 7.3\\
pjmo150421-05 & PJMO 0.6 & CCD & 2015-04-21 & 5.4\\
pjmo150422-03 & PJMO 0.6 & CCD & 2015-04-22 &4.7\\
pjmo150426-03 & PJMO 0.6 & CCD & 2015-04-26 &7.4\\
prom150428-08 & PROMPT 0.4 & CCD & 2015-04-28 &6.5\\
prom150429-03 & PROMPT 0.4 & CCD & 2015-04-29 & 6.6\\
prom150501-03 & PROMPT 0.4 & CCD & 2015-05-01 & 6.2\\
mcao160710-01 & MCAO 0.6 & CCD & 2016-07-10 & 2.0\\
mcao160711-01 & MCAO 0.6 & CCD & 2016-07-11 & 1.5\\
mcao160711-03 & MCAO 0.6 & CCD & 2016-07-11 & 2.0\\
mcao160712-01 & MCAO 0.6 & CCD & 2016-07-12 & 1.2\\
mcao160718-01 & MCAO 0.6 & CCD & 2016-07-18 & 1.2\\
mcao160721-03 & MCAO 0.6 & CCD & 2016-07-21 & 2.0\\
mcao160805-02 & MCAO 0.6 & CCD & 2016-08-05 & 1.2\\
pjmo160802-03 & PJMO 0.6 & CCD & 2016-08-02 & 4.6\\ 
pjmo160804-02 & PJMO 0.6 & CCD & 2016-08-04 & 5.7\\
pjmo160805-02 & PJMO 0.6 & CCD & 2016-08-05 & 5.6\\
pjmo160806-02 & PJMO 0.6 & CCD & 2016-08-06 & 6.0\\
pjmo160808-03 & PJMO 0.6 & CCD & 2016-08-08 & 4.8\\ 
suho160723-19 & Suhora 0.6 & CCD & 2016-07-23 & 5.5\\
tueb160716-21 & Tuebingen 0.8 & CCD & 2016-07-16 & 2.0\\
tueb160718-20 & Tuebingen 0.8 & CCD & 2016-07-18 & 4.5\\
tueb160719-20 & Tuebingen 0.8 & CCD & 2016-07-19 & 6.0\\
tueb160729-22 & Tuebingen 0.8 & CCD & 2016-07-29 & 4.2\\
tueb160730-20 & Tuebingen 0.8 & CCD & 2016-07-30 & 4.0\\
tueb160801-20 & Tuebingen 0.8 & CCD & 2016-08-01 & 2.6\\
tueb160803-20 & Tuebingen 0.8 & CCD & 2016-08-03 & 2.5\\
tueb160807-20 & Tuebingen 0.8 & CCD & 2016-08-07 & 4.9\\
warw160802-20 & Warwick 1.0m & CCD & 2016-08-02 & 5.0\\
warw160803-20 & Warwick 1.0m & CCD & 2016-08-03 & 4.6\\
warw160804-20 & Warwick 1.0m & CCD & 2016-08-04 & 5.0\\

\enddata
\tablecomments{Data from the Warwick 1.0m telescope was obtained during an 
engineering run.}
\end{deluxetable}

\clearpage

\begin{deluxetable}{llllr}
\tablecolumns{4}
\tablewidth{0pc}
\tablecaption{1982-2006 Detected Independent Frequencies
\label{freq1}
}
\tablehead{
\colhead{Frequency} &  \colhead{Period} & \colhead{Amplitude} &  \colhead{Signal/Noise}   &\\
\colhead{\muHz} & \colhead{s} & \colhead{mma} & \colhead{}
}
\startdata

1982 & & & \\
& & & & \\
$1236.483\pm0.07$ & 808.75 & $16.70\pm0.06$ & 9.4\\
$1431.112\pm0.04$ & 698.76 & $30.86\pm0.06$ & 11.4\\
$1613.842\pm0.05$ & 619.65 & $26.12\pm0.06$ & 9.7\\
$1618.845\pm0.06$ & 617.72 & $29.34\pm0.06$ & 10.9\\
$2368.563\pm0.10$ & 422.20 & $12.52\pm0.06$ & 5.8\\
\\
1984 & & & \\
\\
$1124.701\pm3.5$ & 889.13 & $24.7\pm1.1$  & 7.2\\
$1626.474\pm3.5$ & 614.83 & $25.9\pm1.1$ & 6.8\\
\\
1985 & & & \\
\\
$1166.944\pm0.05$ & 856.94 & $6.43\pm0.06$ & 11\\           
$1176.489\pm0.03$ & 849.99 & $35.70\pm0.06$ & 16.0\\
$1614.554\pm0.04$ & 619.37 & $9.15\pm0.06$ & 4.6\\
$2351.762\pm0.04$ & 425.22 & $14.34\pm0.06$ & 12\\
\\
1986 & & & \\
\\
$1081.025\pm0.03$ & 925.05 & $5.1\pm0.4$ & 4.6\\
$1160.678\pm0.03$ & 861.57 & $4.9\pm0.4$ & 4.5  \\
$1223.803\pm0.02$ & 817.12 & $9.3\pm0.4$ & 9.1  \\
$1233.585\pm0.01$ & 810.65 & $14.9\pm0.4$ & 14.5\\
$1385.240\pm0.03$ & 721.90 & $6.7\pm0.4$ & 6.4  \\
$1426.181\pm0.01$ & 701.17 & $14.2\pm0.4$ & 14.3\\
$1525.030\pm0.01$ & 655.72 & $21.3\pm0.4$ & 21.4\\
$1611.691\pm0.01$ & 620.47 & $20.9\pm0.4$ & 20.9\\
$2157.783\pm0.04$ & 463.44 & $2.4\pm0.4$ & 4  \\  
$2165.564\pm0.03$ & 461.77 & $4.8\pm0.4$ & 4\\
$2368.938\pm0.03$ & 422.13 & $5.8\pm0.4$ & 5.6\\
\\
1990 & & & \\
\\
$1112.933\pm0.03$ & 898.53 & $2.31\pm0.32$ & 4.0\\
$1114.180\pm0.03$ & 897.52 & $2.44\pm0.32$ & 4.1\\
$1118.324\pm0.02$ & 894.20 & $5.24\pm0.32$ & 8.9  \\      
$1119.089\pm0.03$ & 893.58 & $2.83\pm0.32$  & 4.8\\
$1224.216\pm0.03$ & 816.83 & $22.17\pm0.32$  & 4.0 \\       
$1233.413\pm0.03$ & 810.76 & $4.80\pm0.32$   & 8.4\\
$1245.395\pm0.03$ & 802.96 & $2.17 \pm0.32$  & 4.0  \\      
$1288.995\pm0.03$ & 775.80 & $3.50\pm0.32$   & 6.3\\
$1291.229\pm0.03$ & 774.46 & $4.31\pm0.32$   & 7.6  \\      
$1295.400\pm0.03$ & 771.96 & $3.27 \pm0.32$  & 5.8\\
$1297.540\pm0.01$ & 770.69 & $14.76\pm0.32$  & 26.1\\
$1304.075\pm0.03$ & 766.83 & $4.67\pm0.32$   & 8.3\\
$1355.447\pm0.03$ & 737.76 & $2.20\pm0.32$   & 4.0  \\      
$1361.728\pm0.03$ & 734.36 & $2.90\pm0.32$   & 5.2\\
$1368.588\pm0.03$ & 730.68 & $3.32\pm0.32$   & 5.9  \\      
$1375.434\pm0.03$ & 727.04 & $3.18\pm0.32$   & 5.7\\
$1421.041\pm0.02$ & 703.71 & $8.22\pm0.32$   & 15.1\\
$1423.704\pm0.03$ & 702.39 & $3.11\pm0.32$   & 5.8\\
$1427.365\pm0.01$ & 700.59 & $19.39\pm0.32$  & 35.8\\
$1428.663\pm0.03$ & 699.96 & $2.61 \pm0.32$  & 4.8\\
$1433.729\pm0.02$ & 697.48 & $7.29\pm0.32$   & 13.5 \\      
$1435.209\pm0.03$ & 696.76 & $3.50\pm0.32$   & 6.5\\
$1512.798\pm0.02$ & 661.03 & $5.57\pm0.32$   & 10.7  \\     
$1518.661\pm0.02$ & 658.47 & $8.34\pm0.32$   & 15.9\\
$1519.372\pm0.02$ & 658.17 & $5.88\pm0.32$   & 11.2  \\     
$1524.924\pm0.02$ & 655.77 & $5.98\pm0.32$   & 11.2\\
$1525.498\pm0.02$ & 655.52 & $6.95\pm0.32$   & 13.0  \\     
$1611.741\pm0.02$ & 620.45 & $6.09\pm0.32$   & 12.3\\
$1617.380\pm0.03$ & 618.28 & $4.69\pm0.32$   & 9.7   \\     
$1623.709\pm0.02$ & 615.87 & $5.04\pm0.32$   & 10.2\\
$2154.009\pm0.03$ & 464.25 & $4.40\pm0.32$   & 11.2 \\      
$2157.765\pm0.03$ & 463.44 & $2.27\pm0.32$   & 5.8\\
$2358.946\pm0.02$ & 423.92 & $5.63\pm0.32$   & 14.1  \\     
$2362.507\pm0.02$ & 423.28 & $5.71\pm0.32$   & 14.3\\
$2366.408\pm0.03$ & 422.58 & $4.59\pm0.32$   & 11.5\\
\\
1991 & & & \\
\\
$1296.087\pm0.13$ & 771.55 & $10.11\pm0.41$ & 17.7\\
$1296.577\pm0.13$ & 771.26 & $18.25\pm0.41$ & 29.8\\
$1308.777\pm0.05$ & 764.07 & $5.02\pm0.41$ & 9.9\\
$1396.945\pm0.05$ & 715.85 & $3.95\pm0.41$ & 8.4\\
$1419.934\pm0.01$ & 704.26 & $29.3\pm0.41$ & 45.6\\
$1423.333\pm0.04$ & 702.58 & $5.23\pm0.41$ & 6.3\\
$1427.062\pm0.04$ & 700.74 & $5.02\pm0.41$ & 7.4\\
$1443.381\pm0.05$ & 692.82 & $4.81\pm0.41$ & 9.7\\
$2150.307\pm0.05$ & 465.05 & $2.36\pm0.41$ & 4.1\\
$2154.389\pm0.05$ & 464.17 & $2.79\pm0.41$ & 4.6\\
$2157.939\pm0.05$ & 463.41 & $2.62\pm0.41$ & 4.4\\
$2370.121\pm0.04$ & 421.92 & $5.01\pm0.41$ & 9.2\\
$2374.211\pm0.05$ & 421.19 & $3.26\pm0.41$ & 5.7\\
$2378.195\pm0.04$ & 420.49 & $5.58\pm0.41$ & 10.7\\
\\
1992 & & & \\
\\
$1035.357\pm0.05$ & 965.85 & $7.35\pm0.31$ & 4.1\\
$1101.952\pm0.05$ & 907.48 & $6.40\pm0.31$ & 3.7\\
$1195.349\pm0.003$ & 836.58 & $6.90\pm0.31$ & 4.1\\
$1233.263\pm0.001$ & 810.86 & $20.79\pm0.31$ & 13.5\\
$1242.854\pm0.003$ & 804.60 & $14.20\pm0.31$ & 8.8\\
$1265.279\pm0.003$ & 790.34 & $12.25\pm0.31$ & 7.1\\
$1420.845\pm0.001$ & 703.81 & $20.59\pm0.31$& 13.1\\
$1438.411\pm0.003$ & 695.21 & $17.71\pm0.31$ & 11.9\\
$1622.795\pm0.003$ & 616.22 & $14.26\pm0.31$ & 9.2\\
$1628.812\pm0.004$ & 613.94 & $10.9\pm0.31$7 & 6.5\\
$2162.104\pm0.007$ & 462.51 & $4.26\pm0.31$ & 4.2\\
$2166.094\pm0.007$ & 461.66 & $5.66\pm0.31$ & 4.8\\
$2351.041\pm0.007$ & 425.34 & $6.75\pm0.31$ & 5.2\\
$2359.162\pm0.007$ & 423.88 & $6.55\pm0.31$& 9.2\\\
$2366.443\pm0.007$ & 422.58 & $7.17\pm0.31$ & 6.3\\
\\
1994 & & & \\
\\
$939.264\pm0.007$ & 1064.66 & $2.36\pm0.10$ & 5.7\\
$1024.773\pm0.011$ & 975.83 & $3.46\pm0.10$ & 8.8\\
$1064.931\pm0.012$ & 939.03 & $2.82\pm0.10$ & 7.4\\
$1106.833\pm0.013$ & 903.48 & $2.56\pm0.10$ & 7.0\\
$1113.548\pm0.012$ & 898.03 & $3.95\pm0.10$ & 9.4\\
$1164.637\pm0.012$ & 858.64 & $3.12\pm0.10$ & 8.6\\
$1176.684\pm0.013$ & 849.85 & $2.79\pm0.10$ & 7.2\\
$1224.306\pm0.012$ & 816.79 & $3.28\pm0.10$ & 9.1\\
$1234.488\pm0.013$ & 810.05 & $2.70\pm0.10$ & 7.7\\
$1235.491\pm0.005$ & 809.39 & $13.13\pm0.10$ & 37.3\\
$1242.357\pm0.013$ & 804.92 & $3.33\pm0.10$ & 9.1\\
$1246.494\pm0.013$ & 802.25 & $2.69\pm0.10$ & 7.8\\
$1286.550\pm0.003$ & 777.27 & $9.45\pm0.10$ & 27.6\\
$1291.023\pm0.009$ & 774.58 & $6.07\pm0.10$& 17.8\\
$1293.244\pm0.009$ & 773.25 & $5.85\pm0.10$ & 16.1\\
$1297.737\pm0.005$ & 770.57 & $21.5\pm0.10$ & 61.9\\
$1298.710\pm0.010$ & 769.99 & $4.45\pm0.10$ & 12.5\\
$1304.464\pm0.009$ & 766.64 & $6.84\pm0.10$ & 19.9\\ 
$1305.352\pm0.013$ & 766.08 & $2.54\pm0.10$ & 7.0\\
$1309.003\pm0.013$ & 763.94 & $2.76\pm0.10$ & 8.4\\
$1312.045\pm0.013$ & 762.17 & $2.67\pm0.10$ & 7.9\\
$1419.641\pm0.003$ & 704.40 & $18.70\pm0.10$ & 54.4\\
$1422.947\pm0.013$ & 702.77 & $2.86\pm0.10$ & 8.2\\
$1426.395\pm0.003$ & 701.07 & $16.05\pm0.10$ & 46.6\\
$1430.851\pm0.005$ & 698.88 & $10.35\pm0.10$ & 30.1\\
$1433.167\pm0.010$ & 697.75 & $4.06\pm0.10$ & 11.6\\
$1437.607\pm0.006$ & 695.60 & $8.28\pm0.10$ & 24.1\\
$1438.523\pm0.011$ & 695.16 & $3.68\pm0.10$ & 10.7\\
$1440.997\pm0.012$ & 693.96 & $2.48\pm0.10$ & 7.2\\
$1441.910\pm0.009$ & 693.52 & $4.09\pm0.10$ & 11.9\\
$1611.351\pm0.009$ & 620.60 & $5.12\pm0.10$ & 14.2\\
$1617.450\pm0.010$ & 618.26 & $3.61\pm0.10$ & 9.5\\
$1618.545\pm0.009$ & 617.84 & $4.04\pm0.10$ & 11.2\\
$1624.624\pm0.009$ & 615.53 & $5.83\pm0.10$ & 16.1\\
$1625.634\pm0.009$ & 615.14 & $4.89\pm0.10$ & 13.9\\
$2150.498\pm0.010$ & 465.01 & $3.22\pm0.10$ & 9.4\\
$2154.130\pm0.010$ & 464.22 & $4.75\pm0.10$ & 13.7\\
$2157.844\pm0.012$ & 463.43 & $2.70\pm0.10$ & 7.6\\
$2358.880\pm0.010$ & 423.93 & $4.53\pm0.10$ & 13.5\\
$2362.636\pm0.005$ & 423.26 & $9.29\pm0.10$ & 31.2\\
$2366.505\pm0.011$ & 422.56 & $4.26\pm0.10$ & 13.2\\
\\
1996 & & & \\
\\
$937.695\pm0.028$ & 1066.44 & $4.27\pm0.28$ & 3.5\\
$1024.276\pm0.028$ & 976.30 & $6.78\pm0.28$ & 6.4\\
$1104.483\pm0.025$ & 905.40 & $5.55\pm0.28$ & 5.1\\
$1178.461\pm0.025$ & 848.56 & $5.34\pm0.28$ & 5.5\\
$1227.936\pm0.022$ & 814.37 & $7.66\pm0.28$ & 7.5\\
$1233.14\pm0.015$  & 810.94 & $13.50\pm0.28$ & 12.1\\
$1234.506\pm0.015$ & 810.04 & $12.53\pm0.28$ & 11.2\\
$1247.959\pm0.022$ & 801.31 & $7.84\pm0.28$ & 4.1\\
$1261.275\pm0.015$ & 792.85 & $11.08\pm0.28$ & 11.6\\
$1291.169\pm0.015$ & 774.49 & $10.71\pm0.28$ & 9.7\\
$1294.899\pm0.025$ & 772.26 & $6.95\pm0.28$ & 7.9\\
$1297.194\pm0.010$ & 770.89 & $22.05\pm0.28$ & 2.4\\
$1420.057\pm0.010$ & 704.20 & $20.26\pm0.28$ & 19.7\\
$1426.443\pm0.010$ & 701.04 & $18.41\pm0.28$ & 8.3\\
$1430.253\pm0.012$ & 699.18 & $13.15\pm0.28$ & 12.8\\
$1436.261\pm0.022$ & 696.25 & $8.20\pm0.28$ & 9.1\\
$1628.296\pm0.028$ & 614.14 & $6.13\pm0.28$ & 6.3\\
$2154.117\pm0.028$ & 464.23 & $7.40\pm0.28$ & 7.1\\
$2358.806\pm0.025$ & 423.94 & $6.01\pm0.28$ & 4.7\\
$2362.617\pm0.014$ & 423.26 & $9.87\pm0.28$ & 9.0\\
$2367.288\pm0.025$ & 422.42 & $8.92\pm0.28$ & 11.6\\
\\\
2000 & & & \\
\\
$938.991\pm0.015$ & 1064.95 & $3.09\pm0.01$ & 9.4\\
$946.238\pm0.015$ & 1056.82 & $1.20\pm0.01$ & 4.0\\
$1110.999\pm0.015$ & 900.09 & $1.98\pm0.01$ & 7.0\\
$1171.564\pm0.015$ & 853.56 & $1.89\pm0.01$ & 7.0\\
$1173.021\pm0.012$ & 852.50 & $2.32\pm0.01$ & 8.8\\
$1251.851\pm0.004$ & 798.82 & $3.21\pm0.01$ & 12.5  \\
$1254.503\pm0.002$ & 797.13 & $8.60\pm0.01$ & 33.1 \\
$1255.583\pm0.002$ & 796.44 & $14.72\pm0.01$ & 67.1\\
$1256.248\pm0.005$ & 796.02 & $8.04\pm0.01$ & 42.6\\
$1257.268\pm0.010$ & 795.38 & $3.05\pm0.01$ & 15.2\\
$1258.288\pm0.010$ & 794.73 & $2.06\pm0.01$ & 5.7\\
$1296.603\pm0.001$ & 771.25 & $28.08\pm0.01$ & 109.5\\ 
$1378.795\pm0.008$ & 725.27 & $4.33\pm0.01$ & 17.3\\
$1379.737\pm0.010$ & 724.78 & $2.64\pm0.01$ & 10.4\\
$1420.101\pm0.001$ & 704.18 & $29.78\pm0.01$ & 118.8\\ 
$1423.597\pm0.010$ & 702.45 & $3.08\pm0.01$ & 12.3\\
$1736.660\pm0.014$ & 575.82 & $1.02\pm0.01$ & 4.0\\
$2150.515\pm0.013$ & 465.01 & $2.99\pm0.01$ & 11.2\\
$2154.040\pm0.013$ & 464.24 & $5.38\pm0.01$ & 19.9\\
$2157.736\pm0.012$ & 463.45 & $2.51\pm0.01$ & 9.5\\
$2359.118\pm0.008$ & 423.89 & $5.51\pm0.01$ & 23.7\\
$2366.271\pm0.010$ & 422.61 & $5.90\pm0.01$ & 25.1\\
\\
2006 & & & \\
\\
$923.976\pm0.001$ & 1082.28 & $1.43\pm0.01$ & 5.2\\
$938.216\pm0.002$ & 1065.85 & $1.23\pm0.01$ & 4.4\\
$1024.496\pm0.002$ & 976.09 & $1.44\pm0.01$& 4.8\\
$1033.760\pm0.002$ & 967.34 & $1.85\pm0.01$ & 6.9\\
$1039.075\pm0.001$ & 962.39 & $7.95\pm0.01$ & 27.2\\
$1039.474\pm0.001$ & 962.02 & $2.84\pm0.01$ & 9.7\\
$1041.535\pm0.001$ & 960.12 & $1.18\pm0.01$ & 4.3\\
$1044.381\pm0.002$ & 957.50 & $1.63\pm0.01$ & 6.2\\
$1113.582\pm0.001$ & 898.02 & $2.71\pm0.01$ & 9.4\\
$1120.404\pm0.001$ & 892.53 & $2.09\pm0.01$ & 7.3\\
$1120.902\pm0.001$ & 892.14 & $2.98\pm0.01$ & 10.2\\
$1121.704\pm0.002$ & 891.50 & $1.26\pm0.01$ & 4.4\\
$1130.144\pm0.002$ & 884.84 & $1.91\pm0.01$ & 7.3\\
$1161.552\pm0.001$ & 860.92 & $2.74\pm0.01$ &  9.6\\
$1173.015\pm0.001$ & 852.50 & $7.26\pm0.01$ & 25.4\\
$1178.096\pm0.002$ & 848.83 & $1.13\pm0.01$ & 4.3\\
$1184.470\pm0.002$ & 844.26 & $1.64\pm0.01$ & 6.2\\
$1222.199\pm0.002$ & 818.20 & $1.72\pm0.01$ & 6.3\\
$1222.751\pm0.001$ & 817.83 & $5.04\pm0.01$ & 18.3\\
$1222.945\pm0.001$ & 817.70 & $4.59 \pm0.01$& 5.1\\
$1228.185\pm0.002$ & 814.21 & $2.71\pm0.01$ & 9.4\\
$1228.791\pm0.001$ & 813.81 & $5.27\pm0.01$ & 19.0\\
$1234.124\pm0.001$ & 810.29 & $24.87\pm0.01$ & 88.0\\
$1239.510\pm0.001$ & 806.77 & $5.07\pm0.01$ & 18.3\\
$1240.237\pm0.002$ & 806.30 & $2.85\pm0.01$ & 9.9\\
$1244.790\pm0.002$ & 803.35 & $1.90\pm0.01$ & 6.9\\
$1245.219\pm0.001$ & 803.07 & $4.75\pm0.01$ & 17.1\\
$1246.032\pm0.001$ & 802.55 & $4.44\pm0.01$ & 15.2\\
$1421.012\pm0.002$ & 703.72 & $2.81\pm0.01$ & 7.1\\
$1429.209\pm0.001$ & 699.69 & $5.65\pm0.01$ & 22.5\\
$1512.141\pm0.002$ & 661.31 & $1.79\pm0.01$ & 6.4\\
$1736.301\pm0.001$ & 575.94 & $16.38\pm0.01$ & 75.2\\
$1737.962\pm0.002$ & 575.39 & $1.80\pm0.01$ & 7.6\\
$1741.666\pm0.001$ & 574.16 & $11.01\pm0.01$ & 49.7\\
$1743.733\pm0.001$ & 573.48 & $5.59\pm0.01$ & 8.4\\
$1749.083\pm0.001$ & 571.73 & $11.88\pm0.01$ & 50.2\\
$1856.845\pm0.002$ & 538.55 & $1.44\pm0.01$ & 6.4\\
$2150.393\pm0.001$ & 465.03 & $4.06\pm0.01$ & 17.3\\
$2154.223\pm0.001$ & 464.20 & $5.50\pm0.01$ & 21.8\\
$2158.073\pm0.001$ & 463.38 & $7.24\pm0.01$ & 29.0\\
$2359.052\pm0.001$ & 423.90 & $5.94\pm0.01$ & 22.1\\
$2363.058\pm0.002$ & 423.18 & $1.7\pm0.01$1 & 6.0\\
$2366.524\pm0.001$ & 422.56 & $6.31\pm0.01$ & 23.0\\

\enddata
\end{deluxetable}
\clearpage
\begin{deluxetable}{llllr}
\tablecolumns{4}
\tablewidth{0pc}
\tablecaption{2007-2016 Detected Independent Frequencies
\label{freq2}
}
\tablehead{
\colhead{Frequency} &  \colhead{Period} & \colhead{Amplitude} &  \colhead{Signal/Noise}   &\\
\colhead{\muHz} & \colhead{s} & \colhead{mma} & \colhead{} 
}
\startdata

2007 & & & \\
\\
$1088.951\pm0.05$ &918.31&$6.9\pm0.7$ & 5\\
$1121.169\pm0.04$&891.93&$10.19\pm0.7$ & 8\\
$1233.956\pm0.02$&810.40&$23\pm0.7$ & 20 \\
$1251.193\pm0.05$&799.24&$9.56\pm0.7$ & 8 \\
$1735.444\pm0.03$&576.22&$13.47\pm0.7$ & 13\\
$2156.288\pm0.03$&463.76&$7.26\pm0.7$ & 7 \\
\\
2008 & & & \\
\\
$1235.745\pm0.004$ & 809.23& $25.14\pm0.33$ & 23.9\\
$1735.716\pm0.004$ & 576.13& $21.77\pm0.33$ &22.6\\
$1741.459\pm0.009$ & 574.23& $10.14\pm0.33$ & 9.8\\
$1750.185\pm0.012$ & 571.37& $4.47\pm0.33$ &5.6\\
$2150.245\pm0.011$ & 465.06& $5.66\pm0.33$ & 7.7\\
$2158.416\pm0.01$ & 463.30&  $11.07\pm0.33$& 11\\
$2358.895\pm0.01$ & 423.93 & $7.95\pm0.33$ &10.2\\
$2366.872\pm0.01$ & 422.50&  $7.93\pm0.33$ &10.2\\
\\
2009 & & & \\
\\
$1088.445\pm0.018$ & 918.74 & $7.82\pm0.28$ &9 \\
$1235.845\pm0.007$ & 809.16 & $9.64\pm0.28$ &22.4\\
$1236.849\pm0.018$ & 808.51 & $7.91\pm0.28$ &9.1\\
$1300.029\pm0.018$ & 576.29 & $4.10\pm0.28$ &4.7\\
$1735.699\pm0.006$ & 576.14 & $20.60\pm0.28$ &22.8\\
$1741.415\pm0.018$ & 574.25 & $7.15\pm0.28$ &8.8\\
$1750.525\pm0.012$ & 571.26 & $5.98\pm0.28$ &6.8\\
$2150.261\pm0.014$ & 465.06 & $5.22\pm0.28$ &6.5\\
$2158.47	4\pm0.003$ & 463.29 & $10.72\pm0.28$ &13.4\\
$2358.952\pm0.026$ & 423.92 & $5.89\pm0.28$ &8.1\\
$2366.893\pm0.018$ & 422.49 & $7.28\pm0.28$ &10.1\\
\\
2010 & & & \\
\\
$1087.79	0\pm0.024$ & 919.29 &$2.47\pm0.15$ & 5.1\\
$1104.397\pm0.023$ & 891.34 &$2.57\pm0.15$ & 5.1\\
$1112.868\pm0.009$ & 898.58 &$6.22\pm0.15$ & 13.2\\
$1121.901\pm0.005$ & 891.34 &$12.6\pm0.15$ & 26.6\\
$1122.674\pm0.021$ & 889.89 &$2.73\pm0.15$ & 5.2\\
$1123.731\pm0.020$ & 891.33 &$3.31\pm0.15$ & 7.1\\
$1200.999\pm0.017$ & 832.64 &$3.33\pm0.15$ & 7.1\\
$1211.906\pm0.021$ & 825.15 &$2.99\pm0.15$ & 6.4\\
$1236.278\pm0.003$ & 808.88 &$18.9\pm0.15$ & 39.7\\
$1237.011\pm0.023$ & 808.41 &$4.10\pm0.15$ & 9.2\\
$1241.721\pm0.014$ & 805.33 &$3.32\pm0.15$ & 7.0\\
$1299.274\pm0.018$ & 769.66 &$3.89\pm0.15$ & 7.6\\ 
$1735.786\pm0.003$ & 576.11 &$21.72\pm0.015$& 46.9\\
$1741.367\pm0.004$ & 574.26 &$8.34\pm0.15$& 17.2\\
$1745.084\pm0.025$ & 573.04 &$2.53\pm0.15$& 5.2\\
$1750.642\pm0.005$ & 571.19 &$10.93\pm0.15$& 25.2\\
$1859.431\pm0.023$ & 537.80 &$2.50\pm0.15$& 5.2\\ 
$2008.645\pm0.025$ & 497.85 &$2.66\pm0.15$& 5.2\\
$2150.138\pm0.025$ & 465.09 &$4.52\pm0.15$& 9.7\\
$2154.432\pm0.025$ & 464.16 &$2.17\pm0.15$& 5.1\\
$2158.524\pm0.006$ & 463.28 &$9.95\pm0.15$& 23.4\\
$2358.818\pm0.008$ & 423.94 &$9.05\pm0.15$& 23.1\\
$2362.603\pm0.017$ & 423.26 &$3.19\pm0.15$& 6.7\\
$2366.99	0\pm0.008$ & 422.48 &$7.40\pm0.15$& 19\\
\\
2011 & & & \\
\\
$954.564\pm0.008$ & 1047.59 & $5.2\pm0.17$ & 10.9\\
$954.922\pm0.02$ & 1047.21 & $4.22\pm0.17$ & 8.7\\
$1113.07\pm0.013$ & 898.45 & $5.72\pm0.17$ & 11.6 \\ 
$1113.641\pm0.02$ & 897.96 & $3.2\pm0.17$ & 6.4\\
$1235.415\pm0.004$ & 809.44 & $9.76\pm0.17$ & 20.1\\
$1362.917\pm0.03$ & 733.72 & $1.66\pm0.17$ & 4.0\\
$1511.537\pm0.03$ & 661.58 & $1.85\pm0.17$ & 4.3\\
$1614.827\pm0.016$ & 619.26 & $4.40\pm0.17$ & 10.5\\
$1735.871\pm0.003$ & 576.08 & $20.3\pm0.17$ & 150.5 \\ 
$1736.14\pm0.037$ & 575.99 & $4.70\pm0.17$ & 11.1\\
$1747.123\pm0.037$ & 572.37 & $3.30\pm0.17$ & 7.9\\
$1750.833\pm0.013$ & 571.16 & $3.77\pm0.17$ & 8.9\\
$1856.872\pm0.03$ & 538.54 & $1.62\pm0.17$ & 3.8\\
$2008.742\pm0.02$ & 497.82 & $3.00\pm0.17$ & 7.4\\
$2150.138\pm0.016$ & 465.09 & $4.98\pm0.17$ & 13.4\\   
$2154.294\pm0.028$ & 464.19 & $3.73\pm0.17$ & 8.9\\
$2158.571\pm0.007$ & 463.27 & $9.10\pm0.17$ & 24.6\\
$2358.707\pm0.009$ & 423.96 & $7.63\pm0.17$ & 27.4  \\ 
$2359.911\pm0.04$ & 423.74 & $1.58\pm0.17$ & 2.47\\
$2366.986\pm$ & 422.48 & $8.9\pm0.17$ & 26.4 \\

\\
2012 & & & \\
\\
$1113.254\pm0.014$ & 898.27 & $4.1\pm0.24$ & 6.2\\
$1161.469\pm0.01$ & 860.98 & $5.21\pm0.24$ & 7.7\\
$1211.371\pm0.006$ & 825.51 & $9.89\pm0.24$ & 14.7\\
$1212.822\pm0.009$ & 824.52 & $6.46\pm0.24$ & 9.6\\
$1213.995\pm0.006$ & 823.73 & $8.66\pm0.24$ & 12.8\\
$1215.915\pm0.01$ & 822.43 & $5.72\pm0.24$ &  8.5\\
$1222.598\pm0.004$ & 817.930 & $11.53\pm0.24$ & 17.2\\
$1226.836\pm0.008$ & 815.11 & $6.89\pm0.24$ & 10.3\\
$1233.355\pm0.004$ & 810.80 & $12.53\pm0.24$ & 18.6\\
$1235.431\pm0.009$ & 809.43 & $6.70\pm0.24$ & 9.9\\
$1246.397\pm0.004$ & 802.31 & $13.35\pm0.24$ & 19.9\\
$1258.482\pm0.009$ & 794.61 & $6.26\pm0.24$ & 9.3\\
$1723.487\pm0.015$ & 580.22 & $4.30\pm0.24$ & 6.1\\
$1734.393\pm0.006$ & 576.57 & $8.54\pm0.24$ & 12.1\\
$1735.975\pm0.004$ & 576.05 & $12.4\pm0.24$ & 17.7\\
$1745.543\pm0.012$ & 572.89 & $4.44\pm0.24$ & 6.3\\
$1747.152\pm0.008$ & 572.36 & $6.74\pm0.24$ & 9.5\\
$1748.895\pm0.009$ & 571.79 & $6.36\pm0.24$ & 9.0\\
$1750.345\pm0.008$ & 571.32 & $6.19\pm0.24$ & 8.9\\
$2150.072\pm0.02$ & 465.10& $2.74\pm0.24$ & 4.0\\
$2155.981\pm0.015$ & 463.83 & $3.68\pm0.24$ & 5.2\\
$2158.563\pm0.006$ & 463.27 & $8.86\pm0.24$ & 12.2\\
$2181.89\pm0.015$ & 458.32 & $3.60\pm0.24$ & 4.9\\
$2355.788\pm0.01$ & 424.47 & $5.48\pm0.24$ & 8.7\\
$2358.721\pm0.01$ & 423.96 & $5.17\pm0.24$ & 8.4\\
$2366.807\pm0.009$ & 422.51 & $5.67\pm0.24$ & 9.4\\
$2372.715\pm0.015$ & 421.46 & $3.81\pm0.24$ & 6.2\\
\\
2013 & & & \\
\\
$1086.678\pm0.007$ & 920.24 & $5.28\pm0.18$ & 6.5\\
$1097.639\pm0.014$ & 911.05 & $5.28\pm0.18$ & 5.4\\
$1123.417\pm0.023$ & 890.14 & $4.46\pm0.18$ & 5.9\\
$1235.981\pm0.005$ & 809.07 & $15.5\pm0.18$ & 15.0\\
$1241.938\pm0.013$ & 805.19 & $7.50\pm0.18$ & 6.4\\
$1362.571\pm0.027$ & 733.91 & $4.74\pm0.18$ & 5.2\\
$1614.872\pm0.007$ & 619.24 & $10.3\pm0.18$ & 10.2\\
$1629.282\pm0.015$ & 613.77 & $4.80\pm0.18$ & 5.3\\
$1736.468\pm0.008$ & 575.88 & $11.1\pm0.18$ & 11.4\\
$1737.008\pm0.008$ & 575.70 & $11.2\pm0.18$ & 11.4\\
$1746.007\pm0.024$ & 572.73 & $5.66\pm0.18$ & 5.7\\
$2158.532\pm0.006$ & 463.28 & $8.32\pm0.18$ & 9.4\\
$2162.432\pm0.014$ & 462.44 & $5.55\pm0.18$ &  6.4\\
$2358.635\pm0.007$ & 423.97 & $7.50\pm0.18$ & 9.2\\
$2364.019\pm0.026$ & 423.013 & $3.20\pm0.18$ & 6.4\\
$2367.037\pm0.014$ & 422.47 & $6.82\pm0.18$ & 8.5\\
\\
2014 & & & \\
\\
$1023.122\pm0.010$ & 977.40 & $2.38\pm0.17$ & 6.1\\
$1113.15\pm0.008$ & 898.35 & $3.53\pm0.17$  & 8.7\\
$1158.561\pm0.014$ & 863.14 & $2.33\pm0.17$ & 6.0\\
$1161.856\pm0.008 $ & 860.69 & $3.33\pm0.17$ & 8.6\\
$1171.782\pm0.010 $ & 853.40 & $2.96\pm0.17$ & 8.0\\
$1172.914\pm0.011 $ & 852.58 & $2.20\pm0.17$ & 5.7\\
$1221.633\pm0.004 $ & 818.58 & $5.19\pm0.17$ & 13.6\\
$1230.125\pm0.008 $ & 812.93 & $3.11\pm0.17$ & 8.1\\
$1234.599\pm0.003 $ & 809.98 & $10.08\pm0.17$ & 26.4\\
$1235.145\pm0.001 $ & 809.62 & $8.63\pm0.17$ & 48.7\\
$1236.228\pm0.004 $ & 808.96 & $5.96\pm0.17$ & 15.6\\
$1237.719\pm0.001 $ & 807.94 & $9.60\pm0.17$ & 25.0\\   
$1248.182\pm0.01 $ & 801.17 & $2.79\pm0.17$ & 7.3\\
$1299.081\pm0.003 $ & 769.77 & $10.58\pm0.17$ & 27.6\\
$1311.786\pm0.008 $ & 762.32 & $3.29\pm0.17$ & 8.5\\
$1312.442\pm0.008 $ & 761.94 & $3.10\pm0.17$ &  8.3\\
$1361.767\pm0.018 $ & 734.34 & $1.61\pm0.17$ & 4.1\\
$1371.106\pm0.014 $ & 729.34 & $1.93\pm0.17$ &  5.0\\
$1429.540\pm0.009$ & 699.53 & $2.97\pm0.17$ & 7.5\\
$1511.465\pm0.006 $ & 661.61 & $4.25\pm0.17$ & 10.2\\
$1520.357\pm0.018 $ & 657.74 & $1.53\pm0.17$ & 3.7\\
$1633.117\pm0.023 $ & 612.33 & $1.19\pm0.17$ & 2.7\\
$1735.955\pm0.001 $ & 576.05 & $18.4\pm0.17$ & 43.1\\
$1739.822\pm0.019 $ & 574.77 & $1.47\pm0.17$ & 3.5\\
$1740.357\pm0.012 $ & 574.59 & $2.22\pm0.17$ & 5.3\\
$1741.406\pm0.019 $ & 574.25 & $1.3\pm0.17$ & 3.3\\
$2003.559\pm0.009 $ & 499.11 & $1.9\pm0.17$ & 4.4\\
$2008.717\pm0.018 $ & 497.83 & $1.53\pm0.17$ & 3.5\\
$2150.280\pm0.010 $ & 465.06 & $2.78\pm0.17$ & 6.7\\
$2154.315\pm0.012 $ & 464.18 & $2.55\pm0.17$ & 6.1\\
$2158.634\pm0.002 $ & 463.26 & $14.41\pm0.17$  & 34.8\\
$2358.801\pm0.010 $ & 423.94 & $6.97\pm0.17$ & 18.2\\
$2362.782\pm0.005 $ & 423.23 & $5.86\pm0.17$ & 15.3\\
$2366.854\pm0.003 $ & 422.50 & $8.24\pm0.17$ & 21.4\\
\\
2015 & & & \\
\\
$1172.340\pm0.024$ & 852.99 & $7.92\pm0.40$ &  9.1\\
$1235.760\pm0.006$ & 809.22& $29.01\pm0.33$ & 35.4\\
$1299.269\pm0.021$ & 769.66 & $8.94\pm0.63$ &  10.6\\
$1736.429\pm0.008$ & 575.89 & $21.68\pm0.33$ & 26.7\\
$2150.059\pm0.060$ & 465.10 & $3.004\pm0.70$ & 4\\
$2154.861\pm0.090$ & 464.07 & $2.06\pm0.73$ & 4\\
$2158.634\pm0.011$ & 463.26 & $17.3\pm0.3$ & 22.9\\
$2358.722\pm0.034$ & 423.96 & $5.501\pm0.7$ & 8.1\\
$2362.884\pm0.023$ & 423.21 & $8.48\pm0.70$ & 12.45\\
$2366.742\pm0.024$ & 422.52 & $8.11\pm0.7$ & 11.9\\
\\
2016 & & & \\
\\
$985.856\pm0.015$ & 1014.35 & $2.9\pm0.3$ & 4.0\\
$1024.527\pm0.01$ &976.06 & $5.0\pm 0.3$ & 5.3\\
$1300.192\pm0.013$ & 769.12 & $3.3\pm0.3$ & 4.0\\
$1360.071\pm0.01 $ & 735.26 & $7.1\pm0.3$ & 8.0\\
$1430.250\pm0.01$ & 699.18 & $5.0\pm0.3$ & 5.7\\
$1511.674\pm0.002$ & 661.52 & $20.5\pm0.3$ & 24.4\\
$1620.449\pm 0.006$ & 617.11& $7.1\pm0.3$ & 9.0\\
$1626.906\pm 0.01$ & 614.66& $4.2\pm0.3$ &5.3\\
$1736.647\pm0.003$ & 575.82& $15.3\pm0.3$ &21.2\\
$2150.461\pm 0.01$ & 465.02& $4.1\pm0.3$ &5.2\\
$2154.493\pm0.016$ & 464.15& $ 2.6\pm0.3$ &4.9\\
$2158.579\pm0.003$ & 463.27& $7.2\pm0.3$ &24.9\\
$2358.772\pm0.006$ & 423.95 & $7.2\pm0.3$ & 13.6\\
$2362.575\pm0.01$ & 423.27& $ 5.0\pm0.3$ &9.4\\
$2366.713\pm0.005$ & 422.53& $8.8\pm0.3$ &16.4\\
\enddata
\end{deluxetable}
\clearpage
\begin{deluxetable}{lllllllll}
\tablecolumns{4}
\tablewidth{0pc}
\tablecaption{Probable Spherical Harmonic Indices from Combination Frequencies
\label{modeampstab}
}
\tablehead{\colhead{Freq.}&\colhead{Period}&\colhead{Amp}&\colhead{Eps}&\colhead{\it{l}}&\colhead{P}&\colhead{\it{m}}&\colhead{P}&\colhead{$\theta$}\\
\colhead{(\muHz)} & \colhead{(s)} & \colhead{(mma)} & \colhead{($\times10^{-7}$)}
}
\startdata

\bf{1990} & & & &&&&\\
$1427.365\pm0.01$&700.59&$19.39\pm0.32$&$1.3$& 1 & 0.98 & 1 & 0.99 &$60\pm5$\\
$1297.540\pm0.01$ & 770.69 & $14.76\pm0.32$& & 1 & 0.99 & -1 & 0.99 & 45,63$\pm5$\\

\bf{1994} & & & &&&&&\\
$1297.737\pm0.005$ & 770.57 & $21.5\pm0.10$&$13$& 1 & 0.99 & 1 & 0.85 & $20\pm2$\\
$1419.641\pm0.003$ & 704.40 & $18.70\pm0.10$& & 1 & 0.99 & -1 & 0.7\\
$1235.491\pm0.005$ & 809.39 & $13.13\pm0.10$& & 1 & 0.99 & 1 & 0.66 \\
$1430.851\pm0.005$ & 698.88 & $10.35\pm0.10$& & 1 & 0.9 & 0 & 0.65 \\

\bf{2006} & & & &&&&&\\
$1234.124\pm0.001$ & 810.29 & $24.87\pm0.01$ & $7.6$& 1 & 0.99 & 0 & 0.99 & $72\pm8$\\
$1736.301\pm0.001$ & 575.94 & $16.38\pm0.01$ & & 1 & 0.99 & 1 & 0.88 \\
$1741.666\pm0.001$ & 574.16 & $11.01\pm0.01$ & & 1,2 & 0.3, 0.7 & ? \\
$1749.083\pm0.001$ & 571.73 & $11.88\pm0.01$ & & 1 & 0.8 & ? \\

\bf{2010} & & & & & & & & \\
$1735.786\pm0.003$ & 576.11 &$21.72\pm0.015$& $33$& 1 & 0.99 & -1 & 0.45 &$45\pm5$\\
$1236.278\pm0.003$ & 808.88 &$18.9\pm0.15$ & & 1 & 0.99 & 1 & 0.85 & \\ 
$1121.901\pm0.005$ & 891.34 &$12.6\pm0.15$ & & 1 & 0.95 & 1 & 0.45 & \\
$1750.642\pm0.005$ & 571.19 &$10.93\pm0.15$& & 1 & 0.5 & -1 & 0.45 & \\

\bf{2011} & & & & & & & & \\
$1735.871\pm0.003$ & 576.08 & $20.3\pm0.17$ & $3.2$& 1 & 0.99 & 1 & 0.75 & $20\pm 10$\\

\bf{2014} &&&&&&&\\
$1735.955\pm0.001 $ & 576.05 & $18.4\pm0.17$ & $7.9$ & 1& 0.99& 0 & 0.8 & $20\pm10$\\
$2158.634\pm0.002 $ & 463.26 & $14.41\pm0.17$&  & 1 & 0.99 & 0 & 0.6 \\
$1299.081\pm0.003 $ & 769.77 & $10.58\pm0.17$ & & 1 & 0.95 & 1 & 0.8 \\
\enddata
\tablecomments{P denotes the probability that, based on detected combination frequencies, the mode has the given value of spherical harmonic 
indices. Values of {m}=$\pm$1 are interchangeable, and have little meaning for modes where rotational splitting is
absent.}
\end{deluxetable}

\clearpage
\begin{deluxetable}{llllll}
\tablecolumns{5}
\tablewidth{0pc}
\tablecaption{List of 15 Periods Used in Fitting GD358 and Corresponding Best Fit Periods.  
\label{gd}
}
\tablehead{
\colhead{$k$} & \colhead{Frequency}& \colhead{Period} & \colhead{Mean Uncertainty} & \colhead{HWHM for Band} & \colhead{Best fit periods}  \\
\colhead{}  & \colhead{(\muHz)} & \colhead{(s)} & \colhead{(s)} & \colhead{(s)} & \colhead{(s)} 
}
\startdata
8  &2363.318 & 423.13 & 0.04 & 0.75 & 423.12 \\
9  &2155.544 & 463.92 & 0.04 & 0.9 & 463.87 \\
10 &2007.628 & 497.83 & 0.2 & &  493.14 \\
11 &1857.716 & 538.30 & 0.3 & & 540.80 \\
12 &1741.505 & 574.22 & 0.1 & 2.2 & 574.90 \\
13 &1619.700 & 617.40 & 0.2 & 3.4 & 615.98 \\
14 &1518.160 & 658.69 & 0.5 & 3.0 & 656.61 \\
15 &1428.943 & 699.82 & 0.2 & 5.8 & 701.43 \\
16 &1369.336 & 730.28 & 0.8 & 8.0 & 741.67 \\
17 &1299.147 & 769.74 & 0.2 & 3.8 & 768.84 \\
18 &1238.129 & 807.67 & 0.2 & 17.0 & 809.40 \\
19 &1170.207 & 854.55 & 0.6 & 2.5 & 854.01 \\
20 &1109.271 & 901.49 & 0.8 & 1.9 & 893.00 \\
22 &1032.967 & 968.09 & 1.4 & & 967.71 \\
24 &941.333  & 1062.32 & 3.1 & & 1051.16\\
\hline
   &         &         & $\sigma_{\rm{rms}}$   & 2.318~s\\
   &   		 &		   & BIC (${\rm n_{obs}}=15$, ${\rm n_{par}}=6$)                 & 26.309~s \\
\tableline
\enddata
\end{deluxetable}
\clearpage

\begin{deluxetable}{llllllll}
\tablecolumns{8}
\tablewidth{0pc}
\tablecaption{Grid parameters and Best Fit Parameters 
\label{fitt1}
}
\tablehead{
\colhead{} & \colhead{\teff [K]}& \colhead{Mass [\msun]}& \colhead{\menv} & \colhead{\mhe} & \colhead{\xo} & \qfm  \\
}
\startdata
\multicolumn{8}{c}{Initial Grid} \\
Minimum  & 21{,}000 & 0.500 & $-2.00$ & $-4.00$ & 0.10  & 0.10               &                       \\
Maximum  & 30{,}000 & 0.700 & $-3.40$ & $-7.00$ & 1.00  & 0.80               &                       \\
Step size & 200     & 0.010 & 0.20   & 0.20   & 0.10  & 0.05               &                       \\
Best fit  & 23{,}600    & 0.57    & $-2.0$ & $-5.6$   & 0.50   & 0.20   &  2.399 s        \\
\multicolumn{8}{c}{Refined Grid} \\
Minimum  & \multicolumn{2}{c}{See Fig. \ref{ffit2}} & $-2.00$ & $-5.30$ & 0.30  & 0.18              &                       \\
Maximum  & \multicolumn{2}{c}{See Fig. \ref{ffit2}} & $-2.50$ & $-5.60$ & 0.60  & 0.22               &                       \\
Step size & 50     & N/A & 0.10   & 0.10   & 0.10  & 0.005         &        \\  
Best fit	 & 23{,}650.0	& 0.5706  & $-2.0$ & $-5.5$   & 0.500  & 0.195  & 2.318 s       \\
\enddata
\end{deluxetable}
\clearpage
\begin{deluxetable}{lcccccccc}
\rotate
\tablecolumns{6}
\tabletypesize{\small}
\tablewidth{0pc}
\tablecaption{
\label{comparefits}
A summary of the results of a number of studies of GD358.
}
\tablehead{
\colhead{Reference} & \colhead{${\rm M_*/M_\odot}$} & \colhead{${\rm T_{eff}}$} & 
\colhead{\menv} & \colhead{\mhe} & distance & O frac & \qfm & ${\rm M_c/M_*}$ \\
\colhead{} & \colhead{} & \colhead{K} & \colhead{} & \colhead{} & \colhead{(pc)} & ${\rm M_o/M_*}$ & & \\
}
\startdata
\citet{Bradley94a}      &$0.61\pm 0.03$  &$24,000\pm 1000$ & N.A.    & $5.70^{+0.18}_{-0.30}$ & $42\pm 3$ & 0.80 & 0.80 & 0.90\\
\citet{Dehner95}      &$0.58$          &$24,210$         & $2.6$   & $6.0$                  & N.A.      & 0.50 &      & 0.997\\
\citet{Metcalfe00}     &$0.605$         &$23,100$         & ($2.74$)& $5.97$                 & N.A.      & 0.80 & 0.50 & 0.90\\
\citet{Metcalfe01}     &$0.65$          &$22,600$         &         & $2.74$                 & N.A.      & 0.84 & 0.49 & 0.95\\
\citet{Fontaine02}      &$0.625\pm 0.036$&$24,800\pm 580$  &$2.97\pm 0.21$& $5.80\pm 0.37$  & $43.0\pm 3.2$ & N.A.& N.A.& N.A.\\
\citet{Metcalfe03}       &$0.66$          &$22,900$         &         & $2.00$                 & N.A.      & 0.67 & 0.48 & 0.98\\
this work &$0.571\pm 0.001$&$23,650\pm 600$  &$2.0\pm 0.1$&$5.5\pm 0.5$ & $44.5^{+8.0}_{-3.0}$& 0.50&0.195& 0.90\\
\citet{Nitta12}    &$0.506\pm 0.04$ &$23,740\pm 92$   &         &                        & N.A.      & N.A. & N.A. & N.A.\\
\citet{Bedard17}    & $0.57\pm 0.03$ &$24,940\pm 1020$ &         &                        & $36.6\pm 4.5$& N.A.&N.A.& N.A.\\
\enddata
\tablecomments{The last two rows are spectroscopic results. ${\rm M_c/M_*}$ is the mass coordinate at which point the oxygen abundance drops to zero.}
\end{deluxetable} 
\clearpage


\end{document}